\documentclass[%
 aip,groupedaddress,
 amsmath,amssymb,
 reprint,%
]{revtex4-2}

\pdfoutput=1

\usepackage[T1]{fontenc}
\usepackage[utf8]{inputenc}
\usepackage[english]{babel}
\usepackage{lmodern}

\usepackage{amsfonts}
\usepackage{amssymb}
\usepackage{amsmath}
\usepackage{bm}
\usepackage{bbm}
\usepackage{cases}
\usepackage[english]{babel}
\usepackage{xcolor}
\definecolor{navyblue}{rgb}{0.0, 0.0, 0.5}
\usepackage{latexsym}
\usepackage{booktabs}
\usepackage[protrusion=true, expansion=true]{microtype}
\usepackage[colorlinks=true, breaklinks=false, linkcolor=navyblue, urlcolor=navyblue, citecolor=navyblue]{hyperref}
\usepackage{float}
\usepackage{tabularx}
\usepackage[utf8]{inputenc}
\usepackage[final]{changes}

\usepackage{mathtools}
\usepackage{braket}
\usepackage{siunitx}
\DeclareUnicodeCharacter{2212}{-} 
\DeclareUnicodeCharacter{3B1}{\ensuremath{\alpha}}

\usepackage{graphicx}
\usepackage{dcolumn}
\usepackage{bm}

\usepackage[utf8]{inputenc}
\usepackage[T1]{fontenc}
\usepackage{mathptmx}
\usepackage{etoolbox}
\makeatletter
\def\@email#1#2{%
 \endgroup
 \patchcmd{\titleblock@produce}
  {\frontmatter@RRAPformat}
  {\frontmatter@RRAPformat{\produce@RRAP{*#1\href{mailto:#2}{#2}}}\frontmatter@RRAPformat}
  {}{}
}%
\makeatother
\begin{document}

\preprint{AIP/123-QED}

\title{Nonlinear spin dynamics across Néel phase transition in ferromagnetic/antiferromagnetic multilayers}

\author{O. Busel}
\author{D. Polishchuk}
\author{A. Kravets}
\author{V. Korenivski*}
\email{vk@kth.se}
\affiliation{Nanostructure Physics, Royal Institute of Technology, 10691 Stockholm, Sweden}
\begin{abstract}
We observe strongly nonlinear spin dynamics in ferro-/antiferro-magnetic multilayers, controlled by the number of bilayers in the system, layer thicknesses, as well as temperature, peaking in magnitude near the Néel point of the antiferromagnetic layers just above room temperature. Well above the Néel transition, the individual ferromagnetic layers are exchange decoupled and resonate independently. As the temperature is lowered toward the Néel point, the ferromagnetic proximity effect through the thin antiferromagnetic spacers transforms the system into a weakly coupled macrospin chain along the film normal, which exhibits pronounced standing spin-wave resonance modes, comparable in intensity to the uniform resonance in the ferromagnetic layers. These findings are supported by our micromagnetic simulations showing clear spin-wave profiles with precessional phase lag along the macrospin chain. Well below the Néel transition, the FeMn layers order strongly antiferromagnetically and exchange-pin the ferromagnetic layers to effectively make the multilayer one macrospin. The appearance and intensity of the high-frequency spin-wave modes can thus be conveniently controlled by thermal gating the multilayer. The nonlinearity in the microwave response of the demonstrated material can approach 100\%, large compared to nonlinear materials used in e.g. optics, with second-harmonic generation often at the single percentage level. 
\end{abstract}

\maketitle

\section{Introduction}
Antiferromagnets (AFs) are of growing interest for applications in spin-thermionic devices~\cite{Baltz2018, Jungwirth2018, Jungwirth2016, Wadley2016, Chumak2015, Yang2024, Prokopenko2024IEEE, Jungwirth2018, Wu2022, Azovtsev2024, Han2024, Zhou2022}, operating in the technologically attractive temperature range around the AF's Néel temperature, $T_\mathrm{N}$, typically just above room temperature (RT). Although AF-based devices operating at sub-THz frequencies are still far from widespread use~\cite{Prokopenko2024, Yang2024, To2024, Prokopenko2024IEEE, Jungwirth2018}, a combination of ferromagnetic (F) and AF materials has the potential to close the so-called GHz-THz gap~\cite{Kanistras2024, Wu2022, Saito2023, Gorbatova2024}. Standing spin waves (SSW) are higher-order resonance modes found in ferromagnets at higher frequencies than the uniform ferromagnetic resonance (FMR) and can reach the antiferromagnetic resonance (AFMR) frequency range of $0.1 - 1$~THz~\cite{Zhuang2024, Azovtsev2024, Han2024}. Furthermore, employing AF spacers as mediators of ferromagnetic interlayer exchange in multilayers~\cite{Zhou2022, Grunberg1986, ChengPRL2018, ChengPRB2018, Hagelschuer2016, Takei2015} can extend the material's functionality with thermal control. 

We recently demonstrated~\cite{Polishchuk2021, Polishchuk2023} thermal gating of AF-mediated interlayer exchange in F/AF/F trilayers, which can be used for thermally assisted parallel-to-antiparallel switching in spin-valve devices. A combination of the AF finite-size~\cite{ChengPRL2018, Polishchuk2023, Tobia2008}, F proximity~\cite{Lenz2007, Golosovsky2009}, and AF-exchange bias~\cite{Antel1999} effects was used. The three layers, and especially the thin AF spacer, were designed to achieve ferromagnetic F-F coupling by magnon exchange via the AF spacer when the AF order is soft enough near its $T_\mathrm{N}$ for magnons in the system to F-AF hybridize~\cite{Liu2003,Gomonay2018,Li2020}.

In this paper, we investigate F/AF multilayers with up to eight F layers that can, in a specific thermal regime, show pronounced standing spin wave (SSW) modes in addition to uniform FMR. We optimize the interplay of four surface/interface effects of finite-size, F-proximity, AF-exchange bias, and interlayer exchange in the multilayered material to achieve an artificial \emph{macrospin chain} of the F-layer moments weakly coupled by AF-mediated exchange, where the latter is strongly geometry- and temperature-dependent. The observed strong SSW nonlinearity in the microwave response of the material is promising for spintronic and magnonic applications, such as GHz oscillators and spin-wave logic.

\section{Experimental methods}

\begin{figure}[t]
\centering
\includegraphics[width=1\columnwidth]{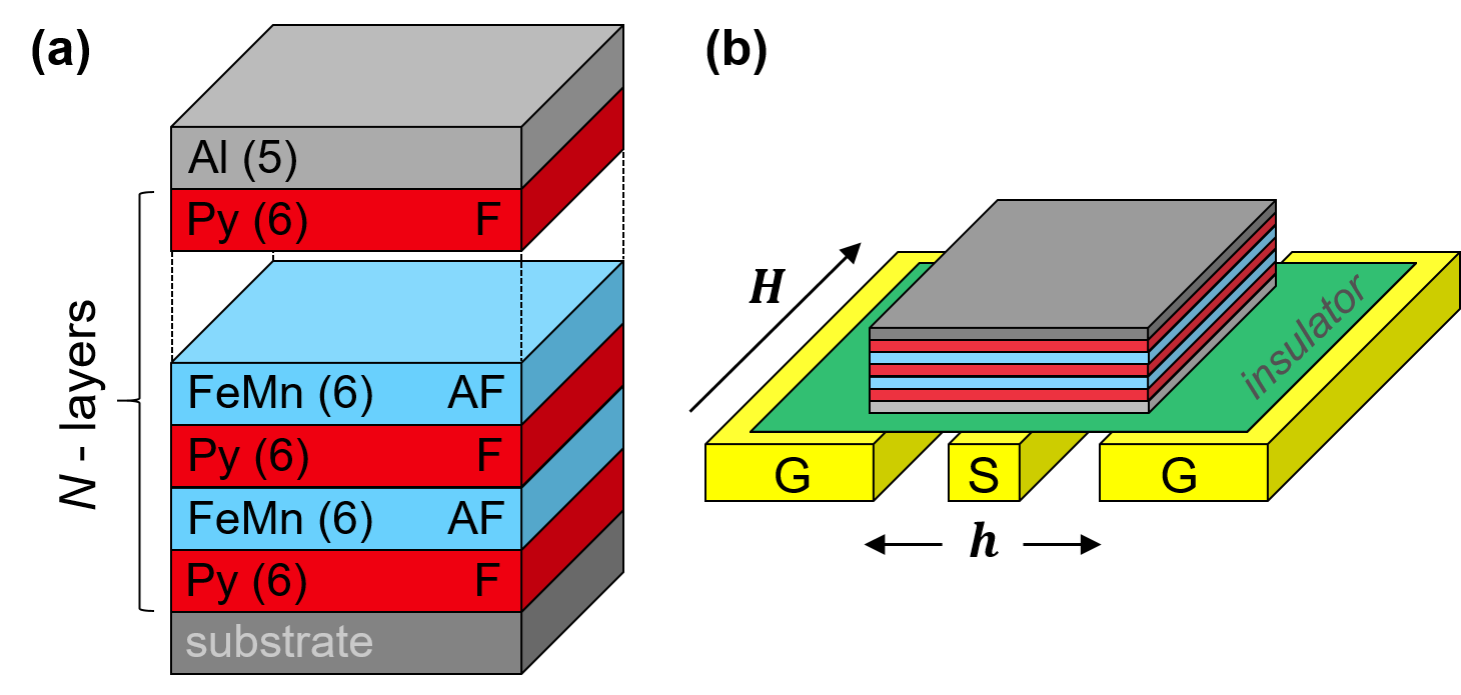}
\caption{(a) Layout of studied F/AF multilayers, with alternating F and AF layers and total number of F layers $N=2-8$; individual layer thicknesses are given in parentheses in nm. (b) Schematic of broadband FMR setup based on coplanar waveguide, with signal (S) and ground (G) lines as well as magnetic field geometry shown, operated in PPMS system at temperatures 2 to 400 K, magnetic fields up to 9 T, and frequencies up to 13 GHz.}
\label{fig:model}
\end{figure}

The layout of the studied F/AF multilayers is shown in Fig.~\ref{fig:model}(a) and has the stack of $N=2$ to 8 F layers (Py or Fe$_{81}$Ni$_{19}$) spaced by $(N-1)$ AF layers (FeMn or Fe$_{50}$Mn$_{50}$). A thin Al capping layer was deposited onto the top surface to protect the multilayers from oxidation. The samples [Py(6)/FeMn(6)]$_{(N-1)}$/Py(6)/Al(5) were deposited at RT on thermally oxidized Si (100) substrates in a magnetic field using DC magnetron sputtering; only the Al layer was deposited using RF sputtering, without breaking vacuum (multi-source DC/RF magnetron UHV Orion system by AJA). The numbers in parentheses denote the respective layer thicknesses in nm. The substrates were pre-etched prior to deposition in a 5~mTorr argon (Ar) atmosphere at 44~W RF-bias power for 5~minutes (working Ar pressure was 3~mTorr and base pressure $\approx 2 \times 10^{-8}$~Torr). After deposition, the samples were annealed in a magnetic field at 100$^{\circ}$C for 1.5~hours and left for overnight cooling to set in unidirectional anisotropy of the F layers (exchange pinning by AF). The AF layers were 6~nm in thickness, which we earlier found to be optimal for maximizing the AF-mediated interlayer exchange~\cite{Polishchuk2021, Polishchuk2023} (near $T_\mathrm{N}$) and thus for enhancing the expected SSW modes in the system.

\section{Results and discussion}

\begin{figure}[t]
\includegraphics[width=1\columnwidth]{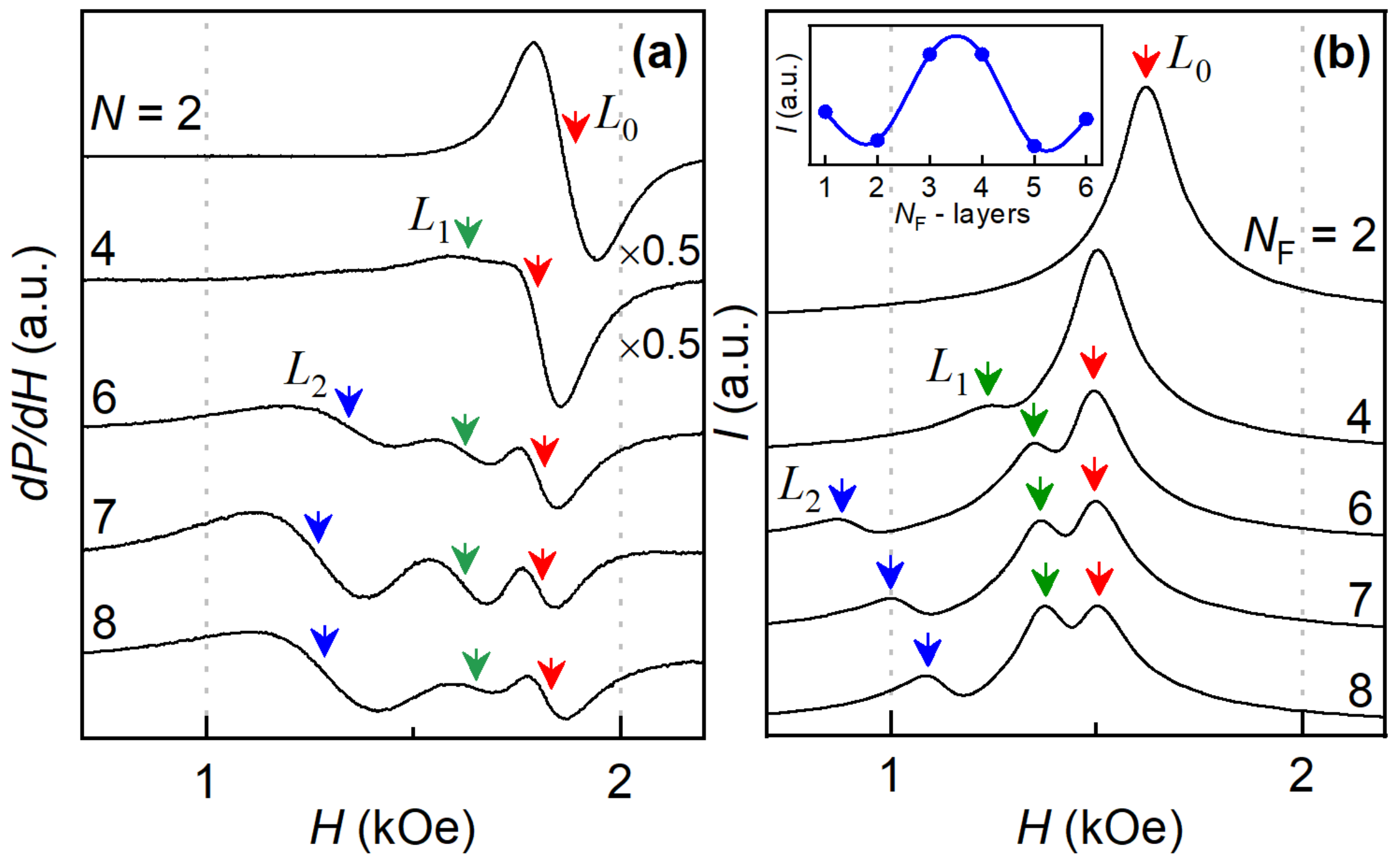}
\caption{(a) FMR spectra measured at RT and 13~GHz for structures with different numbers of F-layers, $N=2,4,6,7,8$, show 3 resonance modes. (b) Micromagnetic simulations for same structures show 3 modes identified as uniform $L_0$ and $L_1$ (split for inner/outer F layers), and SSW $L_2$ pronounced for high $N$; inset shows typical oscillatory profile of SSW mode intensity along \emph{macrospin chain} (normal to film plane, periodic b.c. used in $x-y$ plane, illustrated for $N=6$).}
\label{fig:exp+sim}
\end{figure}

The broadband FMR setup illustrated in Fig.~\ref{fig:model}(b) was used to measure the in-plane  FMR spectra -- the field-derivative of the microwave absorption $dP/dH$ versus the in-plane DC magnetic field $H$ at fixed frequency $f$ and temperature $T$, such as shown in Fig.~\ref{fig:exp+sim}(a) for 13 GHz and RT. The film samples were flip-chipped onto the insulated coplanar waveguide such that the magnetic field $H$ was swept along the direction of the exchange pinning set in during fabrication. The microwave excitation field $h\lesssim1$~Oe was perpendicular to $H$.

Three resonance modes are distinguishable in the experimental FMR spectra of Fig.~\ref{fig:exp+sim}(a): $L_0$ indicated with red, $L_1$ green, and $L_2$ blue arrows. For the F/AF/F trilayer ($N=2$; previously studied system~\cite{Polishchuk2021}), only one pronounced resonance peak is visible. This uniform FMR peak has a weak satellite peak for $N=4$ and is split into two peaks while being strongly diminished in intensity for larger $N$. Especially for $N=7,8$, the multilayer's microwave response is dominated by the emergent $L_2$ mode, with both $L_0$ and $L_1$ significantly smaller in amplitude.

We modeled the behavior of the F/AF multilayers using the micromagnetic simulation package MuMax3~\cite{mumax}, approximating the interlayer exchange coupling via AF and the exchange pinning at the F/AF interfaces by appropriate phenomenological interface/surface anisotropy terms (inner/outer boundary conditions), for all stacks. The results shown in Fig.~\ref{fig:exp+sim}(b) clearly associate the first two resonance peaks ($L_0$ and $L_1$) with the uniform FMR in the system, which is split between the inner and outer F layers since these have 2 and 1 AF interfaces, respectively, resulting in the inner and outer F layers experiencing different effective anisotropy below $T_\mathrm{N}$ (in fact, below the blocking temperature $T_\mathrm{B}$, where the exchange pinning from AF is non-negligible). The inset to Fig.~\ref{fig:exp+sim}(b) shows the magnetization component along the excitation field $h$ of each F layer in the $N=6$ stack. The profile is clearly oscillatory, spin-wave-like, with 3/2 periods edge-to-edge along the thickness, and recreates well the position and dependence on $N$ found in the measured FMR spectra. We were thus able to reproduce in the simulation the overall multilayer spin dynamics as well as that of the individual layers and obtain a good qualitative agreement with the experiment.

Noteworthy in the above micromagnetic analysis is that the detection on the experiment of the higher-order SSW mode (multiple modes, in principle) relies on they inductive effect on the coplanar waveguide, which can be significantly reduced by the oscillating in sign components of the F layers' magnetic moments along the excitation field. The result is that only odd SSW harmonics are likely possible to detect experimentally as the total magnetization and, therefore, stray field from the material for even harmonics fully cancel out. Even for odd harmonics, the pickup signal is reduced by the same effect, with the resonance appearing weaker externally than intrinsically on the individual layers' level. The same considerations are true for the excitation efficiency of odd vs. even SSW modes -- the uniform external excitation field $h$ should have a negligible net coupling to odd SSWs. The major difference between SSWs in ferromagnetic films and those we report herein is that in our case they are mediated by magnon exchange via the AF spacers and therefore are highly tunable by the geometry of the multilayer as well as thermally, conveniently near RT for highly technological AF alloys such as FeMn.

\begin{figure}[t]
\includegraphics[width=0.9\columnwidth]{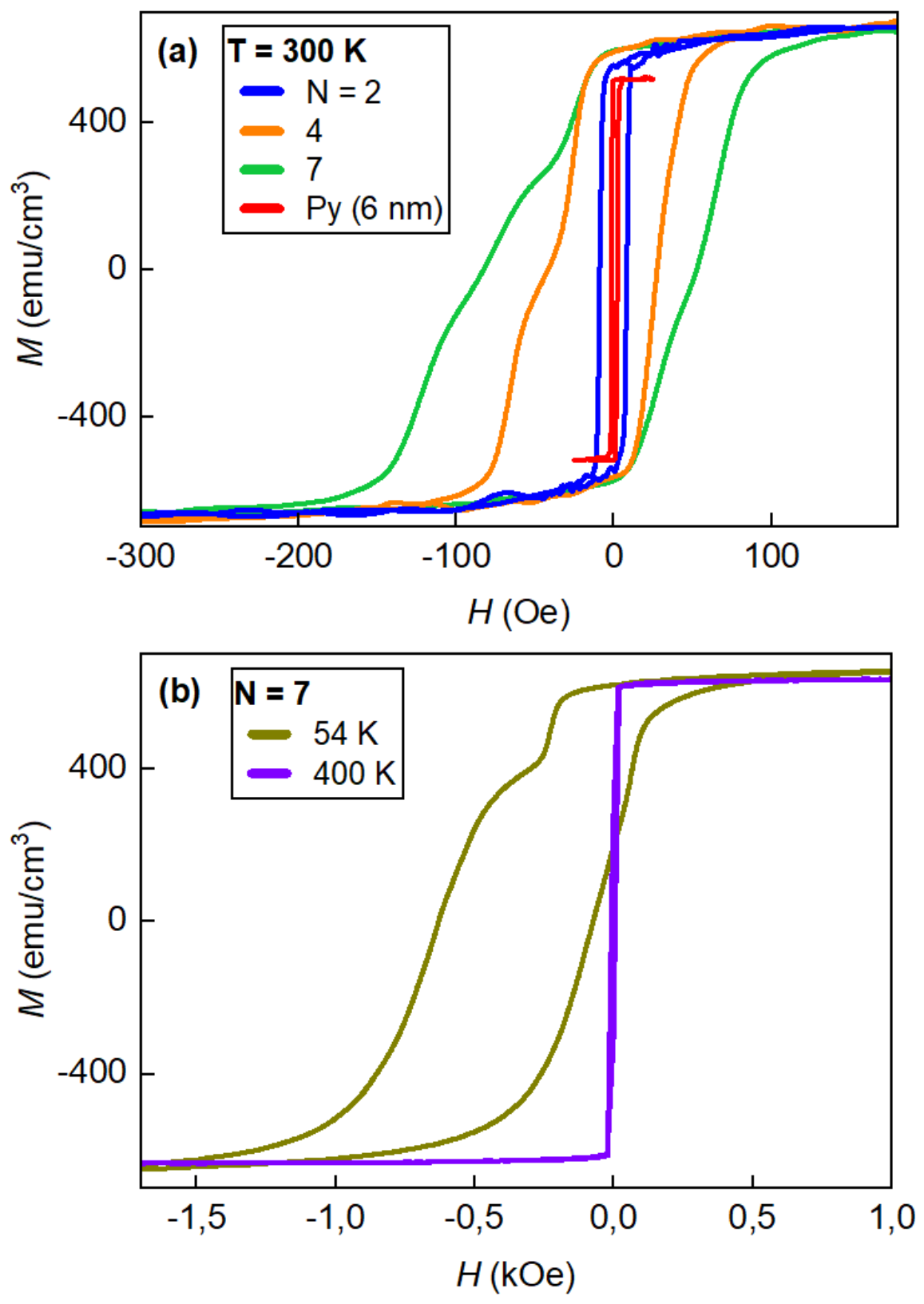}
\caption{(a) VSM magnetization loops for structures with different numbers of F-layers $N$ = 2, 4, and 7 measured at RT. Red loop is for 6~nm single-layer Py film. (b) $M-H$ loops for structure with $N=7$ measured at temperatures well below and well above $T_\mathrm{N}\approx340$~K of AF layers (thin films of FeMn).}
\label{fig:M-H_loops}
\end{figure}

Generally, the presence of magnetic asymmetry enables a more efficient excitation of SSWs. This can be surface magnetic anisotropy (top and/or bottom) different from the interface anisotropy the F layers experience inside the stack. Our vibrating sample magnetometry (VSM) data shown in Fig.~\ref{fig:M-H_loops}(a) do confirm the presence of such magnetic asymmetry in our F/AF multilayers.

The RT VSM magnetization loop for the F/AF/F trilayer sample ($N=2$) shown in blue in Fig.~\ref{fig:M-H_loops}(a) is noticeably wider (higher coercivity) and has a larger magnitude (higher magnetization) compared to that of the reference single Py film shown in red ($M$ in both cases is normalized to respective Py volume). Since the AF layer should nominally have no net magnetization, the saturation magnetization values for these two loops should be expected to coincide. However, the F-proximity effect at the two interfaces in the trilayer induces weak magnetization within the AF layer seen in the data. Of note is that the trilayer loop is essentially symmetric in field at RT$\lesssim T_\mathrm{N}$, which suggests that the single-interface exchange pinning by the AF of the two F layers is too weak to measurably offset the $M-H$ loop at this temperature.

Increasing the number of F layers to $N=4$ and $N=7$ (orange and green loops in Fig.~\ref{fig:M-H_loops}(a)) noticeably offsets the $M-H$ loop, which can now be explained by the one-sided pinning of the top and the bottom F layers and two-sided pinning of the F inner layers,  where the inner layers experience double exchange pinning from their two AF interfaces and are largely responsible for the observed field offset. It should be noted that, in principle, the pinning of the first outer F layer (at bottom/substrate) can differ from that of the second F layer (at top) due to possible differences in the microstructure, interfacial strains, etc. This can be another factor leading to broadening of the main (nominally uniform) FMR in the system.

\begin{figure}[t]
\includegraphics[width=1\columnwidth]{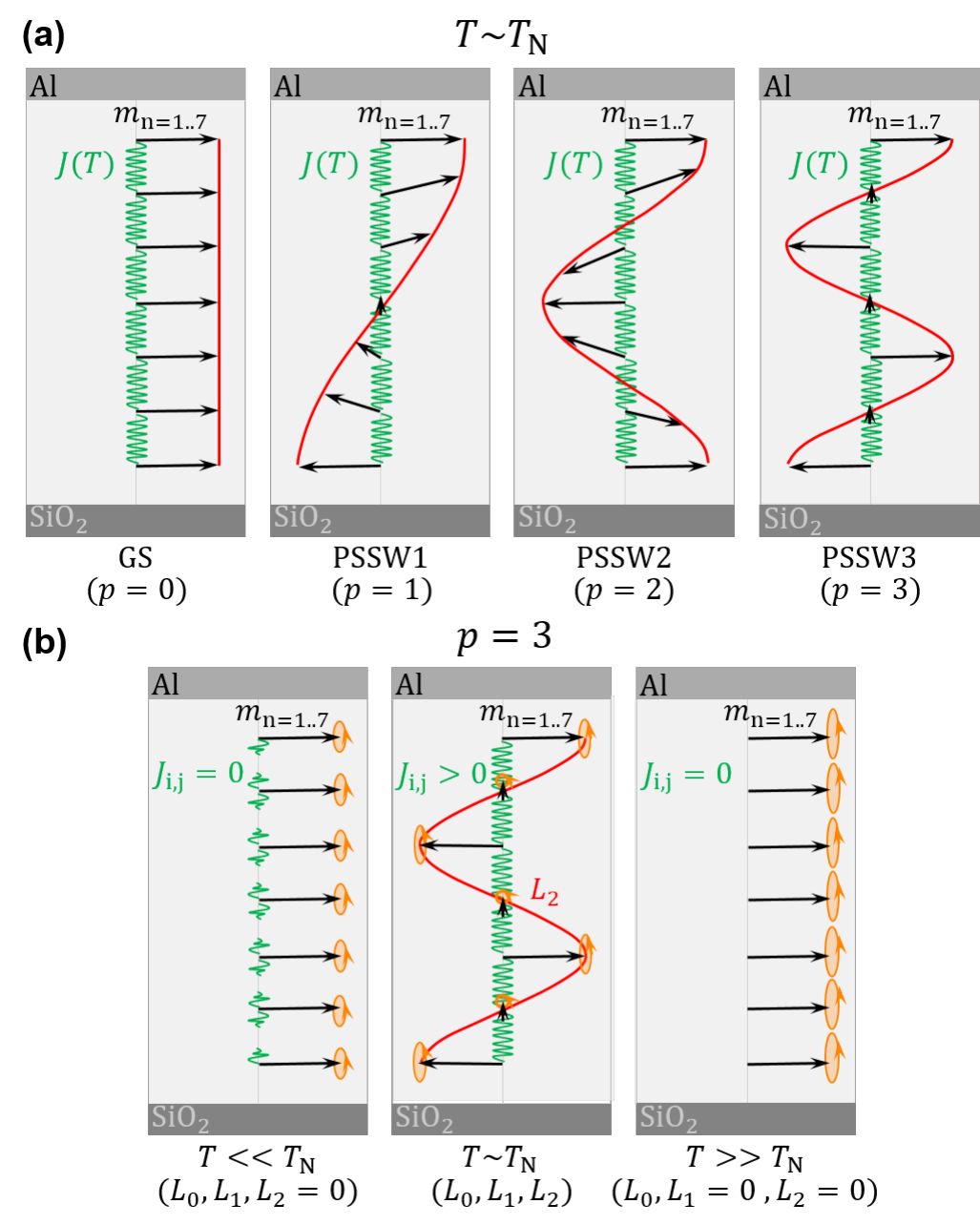}
\caption{(a) Possible PSSW modes in studied system near AF transition, $T\sim T_\text{N}$. (b) $p=3$ mode at temperature well below (left), about (middle), and well above $T_\text{N}$ (right panel).}
\label{fig:PSSW}
\end{figure}

The built-in asymmetry of magnetic anisotropy in our stacks of weakly coupled F layers, inner vs. outer layers, and to a smaller extent bottom vs. top surface, verified by the magnetometry data as discussed above, likely favors excitation of nonuniform spin-wave modes in the system. Indeed, stronger pinning inside a multi-$N$ stack (2 AF interfaces for each F) would be expected to relatively immobilize the inner F layers, while the outer F layers would be more free to respond to the excitation field (situation illustrated in Fig.~\ref{fig:PSSW}; case of no outer surface pinning). Thus, a SSW is enhanced by the AF-induced pinning profile designed into our multilayered material. 

Four key effects interplay in the studied material: finite-size of AF (weaker AF order in thin films), F-proximity (weakly polarizing AF spacers), exchange pinning of F by AF (affecting F anisotropy), and AF-mediated interlayer F-F exchange (near $T_\mathrm{N}$, where the AF magnon gap is suppressed). All four effects are strongly temperature dependent, which is carefully considered below.

The F/AF multilayer with $N_{F}$ = 7 encapsulates all these key effects and exhibits all the resonance modes (uniform and SSW) most clearly. Its $M-H$ loop measured at different temperatures is shown in Fig.~\ref{fig:M-H_loops}(b) and is used to illustrate the physics involved. The effective $T_\text{N}$ of the 6~nm thick FeMn spacer is about 340~K (somewhat below the Néel point of the bulk FeMn material), below which the magnetization of the sample saturates at a slightly high value, consistent with the F-proximity effect partially polarizing the AF layers. It can also be seen in this lower temperature range that the measured net magnetization vs. field of the multilayer consists of two clearly distinguishable minor loops (see e.g. step in $M-H$ for 54~K), which we attribute to the 5 inner (stronger pinned hence stronger offset in $H$, larger step in $M$) and 2 outer F layers (weaker pinned hence weaker offset in $H$, smaller step in $M$). The loop becomes symmetric in field slightly below 340~K (at AF blocking temperature $T_\mathrm{B}$), while signatures of interlayer exchange seen in higher coercivity persist to slightly higher $T$. The AF spacers become fully paramagnetic well above $T_\mathrm{N}$ (fully thermally spin disordered at 400~K), hence providing no pinning and transmit no magnons. As a result, the F layers behave as individual, fully decoupled ferromagnetic films; purple $M-H$ loop measured at 400~K in Fig.~\ref{fig:M-H_loops}(b). Magnetometry thus clearly confirms the presence and main characteristics of the key exchange effects at play in the studied multilayers. 

\begin{figure}[t]
\includegraphics[width=1\columnwidth]{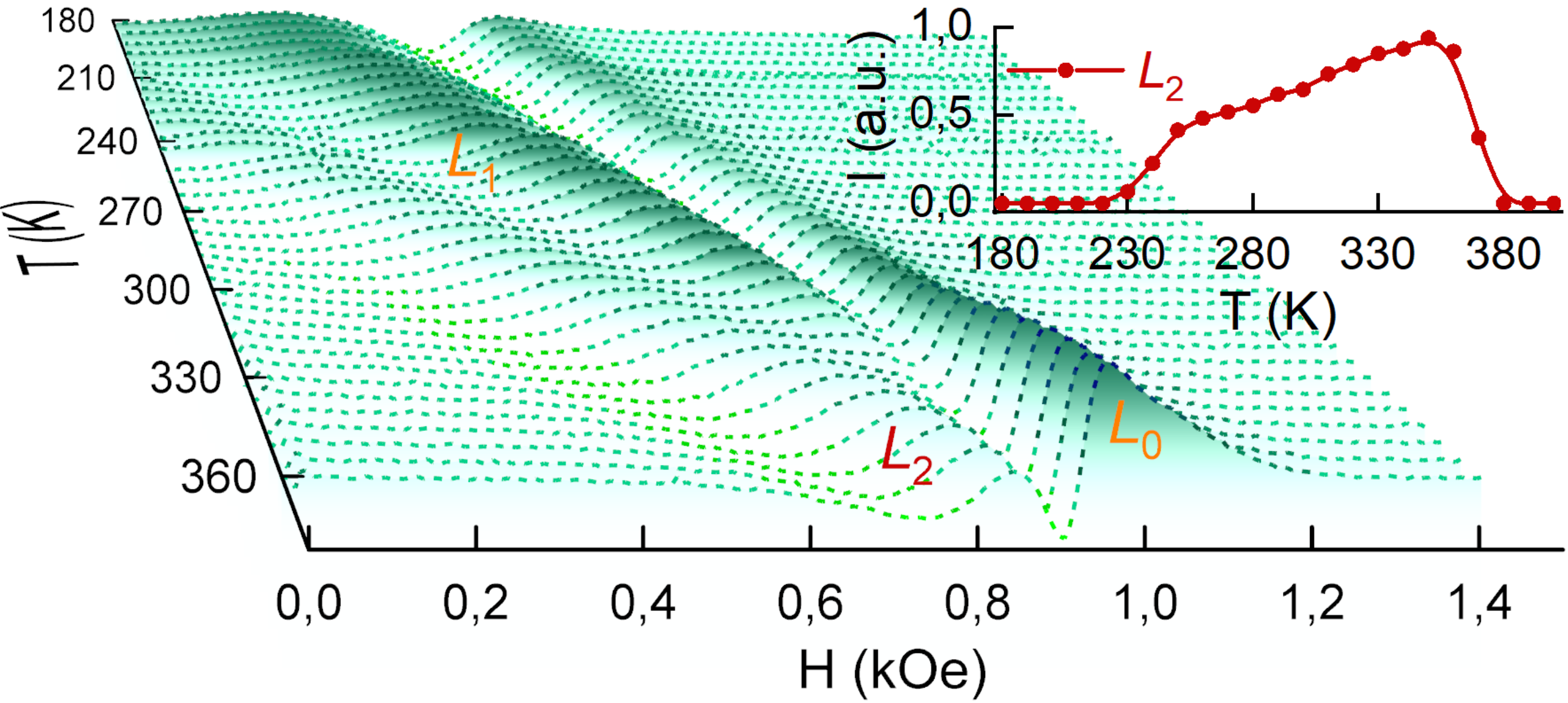}
\caption{FMR spectra shown as field-temperature map for F/AF multilayer with $N=7$, $f=9$~GHz. Inset shows intensity of SSW mode vs. temperature.}
\label{fig:T-colormap}
\end{figure}

Further insight into the rich physics at play in the system across its Néel phase transition is obtained by measuring FMR over the respective temperature range. The results are shown in Fig.~\ref{fig:T-colormap} for $T=180 - 360$~K for the most representative sample with $N=7$. At temperatures well above $T_N$, only one mode is observed, that of uniform FMR where the individual F layers are fully decoupled (paramagnetic AF spacers) and precess in phase in response to the uniform excitation field $h$. 

As the temperature is lowered below 360~K, slightly above $T_\mathrm{N}=340$~K, where F-magnons begin to traverse the AF layers with emerging magnetic order, a collective SSW resonance mode appears, in which the F layers' moments act as macrospins coupled by the relatively weak magnon exchange through the AF spacers. The intensity of this higher-order SSW mode is found to peak at approximately $T_\mathrm{N}$, as shown in the inset to Fig.~\ref{fig:T-colormap}. On further decrease in $T$, the magnon gap in the AF spacers opens up and broadens toward the THz range, so the F-magnons cannot propagate between and couple the F layers in the stack~\cite{Polishchuk2021}. Thus, strong AF ordering at low $T$ quenches the interlayer coupling, and the higher-order SSW mode vanishes, visibly similar to the high-$T$ limit (here, however, due to thermal spin disorder).

\begin{figure}[t]
\includegraphics[width=0.9\columnwidth]{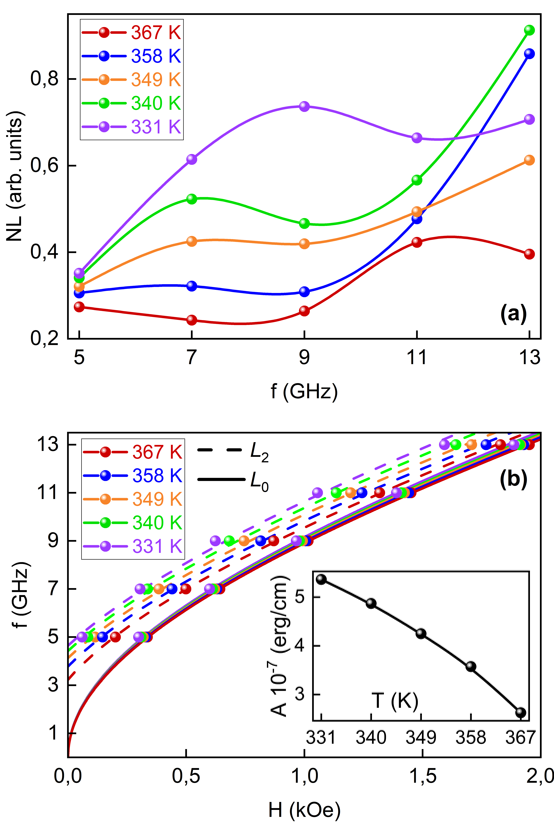}
\caption{(a) Temperature- and frequency-dependent nonlinearity (NL) of FMR spectra for $T=331 \div 367$~K. (b) Kittel-like $f(H)$ resonance relations for $L_0$ (solid) and $L_2$ (dashed) for same temperatures as in (a), where data points are circles and lines are model fits~\cite{Hillebrands2002,Vovk2021}, with extracted exchange stiffness parameter $A(T)$ given in Inset.}
\label{fig:NLandHR}
\end{figure}

The uniform FMR mode is the only resonance present at the highest $T \approx 400$~K. It loses intensity with lowering $T$ and splits into two modes at $T_\mathrm{B} \approx 320$~K, where the asymmetry in anisotropy for inner vs. outer F layers sets in (AF pinning appears at blocking temperature). With further cooling, the second uniform mode $L_1$ overtakes the main uniform mode $L_0$ in intensity, consistent with the larger number of stronger-pinned inner F layers compared to weaker-pinned outer F layer (5 vs. 2) at below $T_\mathrm{B}$.

In a narrow range around $T_N$, the AF layers are thermally agitated, their magnon modes soften and hybridize with the magnons in F~\cite{Polishchuk2021}, such that (weak) interlayer magnon exchange couples the stack ferromagnetically. This weak exchange enables SSW modes, in which the angles between the neighboring F-moments can be large. The effect is observed in a range around $T_\mathrm{N}$, suppressed at low $T$ by strong AF order and at high $T$ by strong spin disorder in the spacers. The rich thermo-magnetic regime demonstrated offers many options for tailoring the properties of the fabricated AF-based material.

All three modes seen in Fig.~\ref{fig:T-colormap} shift toward low fields (high frequencies) in the FMR spectra at low temperature since the system requires less field to resonate with somewhat higher magnetization and higher overall anisotropy. 

We would like to point out, that magnetic phase transitions in multilayered systems where the individual layers have different transition temperatures are never single-Kelvin abrupt; the state-of-the-art in the field, consistent with our previous VSM and FMR results (see~\cite{kravets2012temperature,kravets2015spin,kravets2016anisotropic} and refs. therein), is transition widths of tens of Kelvin (still attractive technologically), which is why the exchange effects extend into the nominally paramagnetic range. Of further note is that $L0$ branches out into $L1$ below $T_\text{B}$, which is why its intensity is lower in this temperature range, while that of $L1$ increases; the combined intensity of $L0$ and $L1$ (uniform FMR of inner and outer Py layers) does not change significantly across $T_\text{B}$.

In Fig.~\ref{fig:NLandHR}(a), we showcase a giant nonlinearity of the microwave response of the material approaching 100\%. This data was obtained by measuring the FMR spectra at several fixed frequencies between 5 and 13 GHz. The nonlinearity parameter NL is taken to be the ratio of the intensity of the SSW mode $L_2$ and the main uniform mode $L_0$ at select temperatures above $T_\mathrm{B}$ (above the splitting of the uniform mode). It increases non-monotonically toward and is largest at 13~GHz, the limitation of our FMR setup. As discussed above, the effects of finite-size, proximity, pinning, and interlayer exchange in the studied multilayer strongly depend on temperature, and so do the resonance frequencies and their field dependence, shown in Fig.~\ref{fig:NLandHR}(b). Even though a close correspondence is not expected for the rather complex system studied herein, we find a surprisingly good fit using Kittel's formalism~\cite{Kittel1958,Hillebrands2002,Vovk2021,Herring1952,Jorzick1999} for $L_0(H)$ and $L_2(H)$: 
\begin{align}
f = \frac{\gamma}{2\pi}\sqrt{\left(H+\frac{2A}{M}\left(\frac{p\pi}{d}\right)^2\right)
{\left(H+\frac{2A}{M}\left(\frac{p\pi}{d}\right)^2+4\pi M\right)}}\notag,
\label{eq:PSSW}
\end{align}
where $\gamma$ is the gyromagnetic ratio, $H$ -- applied magnetic field, $A$ -- exchange stiffness constant, $M$ -- magnetization, $p$ -- SSW order, $d$ -- sample thickness (72~nm in our case). Relatively weak anisotropy in the system (zero to tens of Oe near $T_\textbf{N}$) on the scale of the external field ($\sim 1$~kOe) did not influence the fitting results significantly and is not shown explicitly in the formula above. The fitting yielded $A$, with $M$ determined from the VSM measurements.

The exchange stiffness constant $A$ extracted from the fitting in the vicinity of the Néel transition, in our case representing the AF-mediated interlayer F-F coupling, is shown in the inset to Fig.~\ref{fig:NLandHR}(b) and, as expected, rises in value on lowering $T$ from above $T_\mathrm{N}$. We hope our experimental results will inspire further theoretical work on nonlinear spin dynamics in F/AF multilayered materials.

Fig.~\ref{fig:PSSW} graphically illustrates the physics of the perpendicular SSW (standing spin wave perpendicular to film plane, PSSW) consistent with our measurements, micromagnetic simulations, and SW analytics detailed above. The outer surfaces of the outer Py layers face nonmagnetic SiO$_2$ and Al hence are free from anisotropy, which corresponds to the no-surface-pinning case for PSSW excitation~\cite{Hillebrands2002}. The leftmost panel in (a) shows the uniform mode ($p=0$, for visual clarity no distinction is made between $L_0$ and $L_1$ here, inner vs. outer Py); the rightmost panel is the $p=3$ mode corresponding to three half-wavelengths. The micromagnetic results shown in Fig.~\ref{fig:exp+sim}(b) indicate most pronounced SSW's in the $N=7$ sample and specifically this $p=3$ mode (see Inset for thickness profile). Well above $T_\text{N}$, at 400~K, the spacers are fully paramagnetic, hence the F layers are fully exchange decoupled and precess independently under FMR, showing a single peak on the experiment ($L_0$). Well below $T_\text{N}$, below about 200~K, the AF order in the spacers becomes strong (AF magnon gap grows correspondingly wide), so the F layers become fully decoupled again but exchange-pinned individually and, as a result, precess individually under the higher anisotropy condition (illustrated using a smaller precession cone for low $T$). Near the AF phase transition (AF magnon gap is small and hybridized by F proximity), the finite ferromagnetic exchange-coupling between the F layers in the stack leads to an effective macrospin-chain along the normal to the film plane, which exhibits PSSW. This AF-mediated interlayer exchange is a sensitive function of temperature, which offers a convenient thermal-control mechanism for microwave applications of the material.

\section{Conclusion}

A hybrid material of alternating thin ferromagnetic and antiferromagnetic layers is found to exhibit a highly nonlinear microwave response, which is strongly dependent on temperature, hence can be controlled thereby, in the technologically important range just above room temperature. The relative nonlinearity approaches 100\%, which is giant compared, for example, to popular optical nonlinear materials generating second harmonics at single percentage levels. The highly-nonlinear material properties demonstrated in this work should be interesting for spin-thermionic and thermo-magnonic applications~\cite{Chumak2015, Yuan2023}, where additional functionality can be achieved using thermal control in a narrow range near room temperature.

\begin{acknowledgments}
The authors thank S. Petersen, M. Kulyk, and M. Persson for assistance with measurements and simulations, and acknowledge support from Olle Engkvist Foundation (2020-207-0460) and Swedish Strategic Research Council (SSF UKR24-0002).
\end{acknowledgments}

\section*{Author declarations}
The authors have no conflicts to disclose.

\section*{Data Availability}
The data that support the findings of this study are available from the corresponding author upon reasonable request.

\bibliography{references}

\begin{thebibliography}{41}%
\makeatletter
\providecommand \@ifxundefined [1]{%
 \@ifx{#1\undefined}
}%
\providecommand \@ifnum [1]{%
 \ifnum #1\expandafter \@firstoftwo
 \else \expandafter \@secondoftwo
 \fi
}%
\providecommand \@ifx [1]{%
 \ifx #1\expandafter \@firstoftwo
 \else \expandafter \@secondoftwo
 \fi
}%
\providecommand \natexlab [1]{#1}%
\providecommand \enquote  [1]{``#1''}%
\providecommand \bibnamefont  [1]{#1}%
\providecommand \bibfnamefont [1]{#1}%
\providecommand \citenamefont [1]{#1}%
\providecommand \href@noop [0]{\@secondoftwo}%
\providecommand \href [0]{\begingroup \@sanitize@url \@href}%
\providecommand \@href[1]{\@@startlink{#1}\@@href}%
\providecommand \@@href[1]{\endgroup#1\@@endlink}%
\providecommand \@sanitize@url [0]{\catcode `\\12\catcode `\$12\catcode
  `\&12\catcode `\#12\catcode `\^12\catcode `\_12\catcode `\%12\relax}%
\providecommand \@@startlink[1]{}%
\providecommand \@@endlink[0]{}%
\providecommand \url  [0]{\begingroup\@sanitize@url \@url }%
\providecommand \@url [1]{\endgroup\@href {#1}{\urlprefix }}%
\providecommand \urlprefix  [0]{URL }%
\providecommand \Eprint [0]{\href }%
\providecommand \doibase [0]{https://doi.org/}%
\providecommand \selectlanguage [0]{\@gobble}%
\providecommand \bibinfo  [0]{\@secondoftwo}%
\providecommand \bibfield  [0]{\@secondoftwo}%
\providecommand \translation [1]{[#1]}%
\providecommand \BibitemOpen [0]{}%
\providecommand \bibitemStop [0]{}%
\providecommand \bibitemNoStop [0]{.\EOS\space}%
\providecommand \EOS [0]{\spacefactor3000\relax}%
\providecommand \BibitemShut  [1]{\csname bibitem#1\endcsname}%
\let\auto@bib@innerbib\@empty
\bibitem [{\citenamefont {Baltz}\ \emph {et~al.}(2018)\citenamefont {Baltz},
  \citenamefont {Manchon}, \citenamefont {Tsoi}, \citenamefont {Moriyama},
  \citenamefont {Ono},\ and\ \citenamefont {Tserkovnyak}}]{Baltz2018}%
  \BibitemOpen
  \bibfield  {author} {\bibinfo {author} {\bibfnamefont {V.}~\bibnamefont
  {Baltz}}, \bibinfo {author} {\bibfnamefont {A.}~\bibnamefont {Manchon}},
  \bibinfo {author} {\bibfnamefont {M.}~\bibnamefont {Tsoi}}, \bibinfo {author}
  {\bibfnamefont {T.}~\bibnamefont {Moriyama}}, \bibinfo {author}
  {\bibfnamefont {T.}~\bibnamefont {Ono}},\ and\ \bibinfo {author}
  {\bibfnamefont {Y.}~\bibnamefont {Tserkovnyak}},\ }\bibfield  {title}
  {\enquote {\bibinfo {title} {Antiferromagnetic spintronics},}\ }\href
  {https://doi.org/10.1103/RevModPhys.90.015005} {\bibfield  {journal}
  {\bibinfo  {journal} {Rev. Mod. Phys.}\ }\textbf {\bibinfo {volume} {90}},\
  \bibinfo {pages} {015005} (\bibinfo {year} {2018})}\BibitemShut {NoStop}%
\bibitem [{\citenamefont {Jungwirth}\ \emph {et~al.}(2018)\citenamefont
  {Jungwirth}, \citenamefont {Sinova}, \citenamefont {Manchon}, \citenamefont
  {Marti}, \citenamefont {Wunderlich},\ and\ \citenamefont
  {Felser}}]{Jungwirth2018}%
  \BibitemOpen
  \bibfield  {author} {\bibinfo {author} {\bibfnamefont {T.}~\bibnamefont
  {Jungwirth}}, \bibinfo {author} {\bibfnamefont {J.}~\bibnamefont {Sinova}},
  \bibinfo {author} {\bibfnamefont {A.}~\bibnamefont {Manchon}}, \bibinfo
  {author} {\bibfnamefont {X.}~\bibnamefont {Marti}}, \bibinfo {author}
  {\bibfnamefont {J.}~\bibnamefont {Wunderlich}},\ and\ \bibinfo {author}
  {\bibfnamefont {C.}~\bibnamefont {Felser}},\ }\bibfield  {title} {\enquote
  {\bibinfo {title} {The multiple directions of antiferromagnetic
  spintronics},}\ }\href {https://doi.org/10.1038/s41567-018-0063-6} {\bibfield
   {journal} {\bibinfo  {journal} {Nature Physics}\ }\textbf {\bibinfo {volume}
  {14}} (\bibinfo {year} {2018}),\ 10.1038/s41567-018-0063-6}\BibitemShut
  {NoStop}%
\bibitem [{\citenamefont {Jungwirth}\ \emph {et~al.}(2016)\citenamefont
  {Jungwirth}, \citenamefont {Marti}, \citenamefont {Wadley},\ and\
  \citenamefont {Wunderlich}}]{Jungwirth2016}%
  \BibitemOpen
  \bibfield  {author} {\bibinfo {author} {\bibfnamefont {T.}~\bibnamefont
  {Jungwirth}}, \bibinfo {author} {\bibfnamefont {X.}~\bibnamefont {Marti}},
  \bibinfo {author} {\bibfnamefont {P.}~\bibnamefont {Wadley}},\ and\ \bibinfo
  {author} {\bibfnamefont {J.}~\bibnamefont {Wunderlich}},\ }\bibfield  {title}
  {\enquote {\bibinfo {title} {Antiferromagnetic spintronics},}\ }\href
  {https://doi.org/10.1038/nnano.2016.18} {\bibfield  {journal} {\bibinfo
  {journal} {Nature Nanotechnology}\ }\textbf {\bibinfo {volume} {11}}
  (\bibinfo {year} {2016}),\ 10.1038/nnano.2016.18}\BibitemShut {NoStop}%
\bibitem [{\citenamefont {Wadley}\ \emph {et~al.}(2016)\citenamefont {Wadley},
  \citenamefont {Howells}, \citenamefont {Železný}, \citenamefont {Andrews},
  \citenamefont {Hills}, \citenamefont {Campion}, \citenamefont {Novák},
  \citenamefont {Olejník}, \citenamefont {Maccherozzi}, \citenamefont {Dhesi},
  \citenamefont {Martin}, \citenamefont {Wagner}, \citenamefont {Wunderlich},
  \citenamefont {Freimuth}, \citenamefont {Mokrousov}, \citenamefont {Kuneš},
  \citenamefont {Chauhan}, \citenamefont {Grzybowski}, \citenamefont
  {Rushforth}, \citenamefont {Edmonds}, \citenamefont {Gallagher},\ and\
  \citenamefont {Jungwirth}}]{Wadley2016}%
  \BibitemOpen
  \bibfield  {author} {\bibinfo {author} {\bibfnamefont {P.}~\bibnamefont
  {Wadley}}, \bibinfo {author} {\bibfnamefont {B.}~\bibnamefont {Howells}},
  \bibinfo {author} {\bibfnamefont {J.}~\bibnamefont {Železný}}, \bibinfo
  {author} {\bibfnamefont {C.}~\bibnamefont {Andrews}}, \bibinfo {author}
  {\bibfnamefont {V.}~\bibnamefont {Hills}}, \bibinfo {author} {\bibfnamefont
  {R.~P.}\ \bibnamefont {Campion}}, \bibinfo {author} {\bibfnamefont
  {V.}~\bibnamefont {Novák}}, \bibinfo {author} {\bibfnamefont
  {K.}~\bibnamefont {Olejník}}, \bibinfo {author} {\bibfnamefont
  {F.}~\bibnamefont {Maccherozzi}}, \bibinfo {author} {\bibfnamefont {S.~S.}\
  \bibnamefont {Dhesi}}, \bibinfo {author} {\bibfnamefont {S.~Y.}\ \bibnamefont
  {Martin}}, \bibinfo {author} {\bibfnamefont {T.}~\bibnamefont {Wagner}},
  \bibinfo {author} {\bibfnamefont {J.}~\bibnamefont {Wunderlich}}, \bibinfo
  {author} {\bibfnamefont {F.}~\bibnamefont {Freimuth}}, \bibinfo {author}
  {\bibfnamefont {Y.}~\bibnamefont {Mokrousov}}, \bibinfo {author}
  {\bibfnamefont {J.}~\bibnamefont {Kuneš}}, \bibinfo {author} {\bibfnamefont
  {J.~S.}\ \bibnamefont {Chauhan}}, \bibinfo {author} {\bibfnamefont {M.~J.}\
  \bibnamefont {Grzybowski}}, \bibinfo {author} {\bibfnamefont {A.~W.}\
  \bibnamefont {Rushforth}}, \bibinfo {author} {\bibfnamefont {K.~W.}\
  \bibnamefont {Edmonds}}, \bibinfo {author} {\bibfnamefont {B.~L.}\
  \bibnamefont {Gallagher}},\ and\ \bibinfo {author} {\bibfnamefont
  {T.}~\bibnamefont {Jungwirth}},\ }\bibfield  {title} {\enquote {\bibinfo
  {title} {Electrical switching of an antiferromagnet},}\ }\href
  {https://doi.org/10.1126/science.aab1031} {\bibfield  {journal} {\bibinfo
  {journal} {Science}\ }\textbf {\bibinfo {volume} {351}},\ \bibinfo {pages}
  {587--590} (\bibinfo {year} {2016})}\BibitemShut {NoStop}%
\bibitem [{\citenamefont {Chumak}\ \emph {et~al.}(2015)\citenamefont {Chumak},
  \citenamefont {Vasyuchka}, \citenamefont {Serga},\ and\ \citenamefont
  {Hillebrands}}]{Chumak2015}%
  \BibitemOpen
  \bibfield  {author} {\bibinfo {author} {\bibfnamefont {A.~V.}\ \bibnamefont
  {Chumak}}, \bibinfo {author} {\bibfnamefont {V.~I.}\ \bibnamefont
  {Vasyuchka}}, \bibinfo {author} {\bibfnamefont {A.~A.}\ \bibnamefont
  {Serga}},\ and\ \bibinfo {author} {\bibfnamefont {B.}~\bibnamefont
  {Hillebrands}},\ }\bibfield  {title} {\enquote {\bibinfo {title} {Magnon
  spintronics},}\ }\href {https://doi.org/10.1038/nphys3347} {\bibfield
  {journal} {\bibinfo  {journal} {Nature Physics}\ }\textbf {\bibinfo {volume}
  {11}},\ \bibinfo {pages} {453} (\bibinfo {year} {2015})}\BibitemShut
  {NoStop}%
\bibitem [{\citenamefont {Yang}\ \emph {et~al.}(2024)\citenamefont {Yang},
  \citenamefont {Kim}, \citenamefont {Lee}, \citenamefont {Xu}, \citenamefont
  {Liu}, \citenamefont {Wang}, \citenamefont {Zhao}, \citenamefont {Kumar},\
  and\ \citenamefont {Yang}}]{Yang2024}%
  \BibitemOpen
  \bibfield  {author} {\bibinfo {author} {\bibfnamefont {D.}~\bibnamefont
  {Yang}}, \bibinfo {author} {\bibfnamefont {T.}~\bibnamefont {Kim}}, \bibinfo
  {author} {\bibfnamefont {K.}~\bibnamefont {Lee}}, \bibinfo {author}
  {\bibfnamefont {C.}~\bibnamefont {Xu}}, \bibinfo {author} {\bibfnamefont
  {Y.}~\bibnamefont {Liu}}, \bibinfo {author} {\bibfnamefont {F.}~\bibnamefont
  {Wang}}, \bibinfo {author} {\bibfnamefont {S.}~\bibnamefont {Zhao}}, \bibinfo
  {author} {\bibfnamefont {D.}~\bibnamefont {Kumar}},\ and\ \bibinfo {author}
  {\bibfnamefont {H.}~\bibnamefont {Yang}},\ }\bibfield  {title} {\enquote
  {\bibinfo {title} {Spin-orbit torque manipulation of sub-terahertz magnons in
  antiferromagnetic α-fe2o3},}\ }\href
  {https://doi.org/10.1038/s41467-024-48431-w} {\bibfield  {journal} {\bibinfo
  {journal} {Nature Communications}\ }\textbf {\bibinfo {volume} {15}}
  (\bibinfo {year} {2024}),\ 10.1038/s41467-024-48431-w}\BibitemShut {NoStop}%
\bibitem [{\citenamefont {Prokopenko}\ \emph {et~al.}(2024)\citenamefont
  {Prokopenko}, \citenamefont {Bankowski}, \citenamefont {Prokopenko},\ and\
  \citenamefont {Slavin}}]{Prokopenko2024IEEE}%
  \BibitemOpen
  \bibfield  {author} {\bibinfo {author} {\bibfnamefont {V.~O.}\ \bibnamefont
  {Prokopenko}}, \bibinfo {author} {\bibfnamefont {E.~N.}\ \bibnamefont
  {Bankowski}}, \bibinfo {author} {\bibfnamefont {O.~V.}\ \bibnamefont
  {Prokopenko}},\ and\ \bibinfo {author} {\bibfnamefont {A.~N.}\ \bibnamefont
  {Slavin}},\ }\bibfield  {title} {\enquote {\bibinfo {title} {The impact of
  temperature on the performance of an active terahertz-frequency signal
  detector based on an antiferromagnetic tunnel junction},}\ }\href
  {https://doi.org/10.1109/TMAG.2024.3440189} {\bibfield  {journal} {\bibinfo
  {journal} {IEEE Transactions on Magnetics}\ }\textbf {\bibinfo {volume}
  {60}},\ \bibinfo {pages} {1--7} (\bibinfo {year} {2024})}\BibitemShut
  {NoStop}%
\bibitem [{\citenamefont {Wu}\ \emph {et~al.}(2022)\citenamefont {Wu},
  \citenamefont {Wang}, \citenamefont {Liu}, \citenamefont {Wang},
  \citenamefont {Chen}, \citenamefont {Chen}, \citenamefont {Li}, \citenamefont
  {Han}, \citenamefont {Miao}, \citenamefont {Yu}, \citenamefont {Wan},
  \citenamefont {Zhao},\ and\ \citenamefont {Chen}}]{Wu2022}%
  \BibitemOpen
  \bibfield  {author} {\bibinfo {author} {\bibfnamefont {X.}~\bibnamefont
  {Wu}}, \bibinfo {author} {\bibfnamefont {H.}~\bibnamefont {Wang}}, \bibinfo
  {author} {\bibfnamefont {H.}~\bibnamefont {Liu}}, \bibinfo {author}
  {\bibfnamefont {Y.}~\bibnamefont {Wang}}, \bibinfo {author} {\bibfnamefont
  {X.}~\bibnamefont {Chen}}, \bibinfo {author} {\bibfnamefont {P.}~\bibnamefont
  {Chen}}, \bibinfo {author} {\bibfnamefont {P.}~\bibnamefont {Li}}, \bibinfo
  {author} {\bibfnamefont {X.}~\bibnamefont {Han}}, \bibinfo {author}
  {\bibfnamefont {J.}~\bibnamefont {Miao}}, \bibinfo {author} {\bibfnamefont
  {H.}~\bibnamefont {Yu}}, \bibinfo {author} {\bibfnamefont {C.}~\bibnamefont
  {Wan}}, \bibinfo {author} {\bibfnamefont {J.}~\bibnamefont {Zhao}},\ and\
  \bibinfo {author} {\bibfnamefont {S.}~\bibnamefont {Chen}},\ }\bibfield
  {title} {\enquote {\bibinfo {title} {Antiferromagnetic–ferromagnetic
  heterostructure-based field-free terahertz emitters},}\ }\href
  {https://doi.org/https://doi.org/10.1002/adma.202204373} {\bibfield
  {journal} {\bibinfo  {journal} {Advanced Materials}\ }\textbf {\bibinfo
  {volume} {34}},\ \bibinfo {pages} {2204373} (\bibinfo {year}
  {2022})}\BibitemShut {NoStop}%
\bibitem [{\citenamefont {Azovtsev}\ and\ \citenamefont
  {Pertsev}(2024)}]{Azovtsev2024}%
  \BibitemOpen
  \bibfield  {author} {\bibinfo {author} {\bibfnamefont {A.~V.}\ \bibnamefont
  {Azovtsev}}\ and\ \bibinfo {author} {\bibfnamefont {N.~A.}\ \bibnamefont
  {Pertsev}},\ }\bibfield  {title} {\enquote {\bibinfo {title}
  {Antiferromagnetic standing spin waves generated in
  $\mathrm{N}\mathrm{i}\mathrm{O}$ thin films by short strain pulses},}\ }\href
  {https://doi.org/10.1103/PhysRevB.110.144430} {\bibfield  {journal} {\bibinfo
   {journal} {Phys. Rev. B}\ }\textbf {\bibinfo {volume} {110}},\ \bibinfo
  {pages} {144430} (\bibinfo {year} {2024})}\BibitemShut {NoStop}%
\bibitem [{\citenamefont {Han}, \citenamefont {Wu},\ and\ \citenamefont
  {Zhang}(2024)}]{Han2024}%
  \BibitemOpen
  \bibfield  {author} {\bibinfo {author} {\bibfnamefont {X.}~\bibnamefont
  {Han}}, \bibinfo {author} {\bibfnamefont {H.}~\bibnamefont {Wu}},\ and\
  \bibinfo {author} {\bibfnamefont {T.}~\bibnamefont {Zhang}},\ }\bibfield
  {title} {\enquote {\bibinfo {title} {Magnonics: Materials, physics, and
  devices},}\ }\href {https://doi.org/10.1063/5.0216094} {\bibfield  {journal}
  {\bibinfo  {journal} {Applied Physics Letters}\ }\textbf {\bibinfo {volume}
  {125}},\ \bibinfo {pages} {020501} (\bibinfo {year} {2024})}\BibitemShut
  {NoStop}%
\bibitem [{\citenamefont {Zhou}\ \emph {et~al.}(2022)\citenamefont {Zhou},
  \citenamefont {Liao}, \citenamefont {Guo}, \citenamefont {Bai}, \citenamefont
  {Zhao}, \citenamefont {Wan}, \citenamefont {Huang}, \citenamefont {Han},
  \citenamefont {Qiao}, \citenamefont {You}, \citenamefont {Chen},
  \citenamefont {Chen}, \citenamefont {Zhou}, \citenamefont {Han},
  \citenamefont {Pan},\ and\ \citenamefont {Song}}]{Zhou2022}%
  \BibitemOpen
  \bibfield  {author} {\bibinfo {author} {\bibfnamefont {Y.}~\bibnamefont
  {Zhou}}, \bibinfo {author} {\bibfnamefont {L.}~\bibnamefont {Liao}}, \bibinfo
  {author} {\bibfnamefont {T.}~\bibnamefont {Guo}}, \bibinfo {author}
  {\bibfnamefont {H.}~\bibnamefont {Bai}}, \bibinfo {author} {\bibfnamefont
  {M.}~\bibnamefont {Zhao}}, \bibinfo {author} {\bibfnamefont {C.}~\bibnamefont
  {Wan}}, \bibinfo {author} {\bibfnamefont {L.}~\bibnamefont {Huang}}, \bibinfo
  {author} {\bibfnamefont {L.}~\bibnamefont {Han}}, \bibinfo {author}
  {\bibfnamefont {L.}~\bibnamefont {Qiao}}, \bibinfo {author} {\bibfnamefont
  {Y.}~\bibnamefont {You}}, \bibinfo {author} {\bibfnamefont {C.}~\bibnamefont
  {Chen}}, \bibinfo {author} {\bibfnamefont {R.}~\bibnamefont {Chen}}, \bibinfo
  {author} {\bibfnamefont {Z.}~\bibnamefont {Zhou}}, \bibinfo {author}
  {\bibfnamefont {X.}~\bibnamefont {Han}}, \bibinfo {author} {\bibfnamefont
  {F.}~\bibnamefont {Pan}},\ and\ \bibinfo {author} {\bibfnamefont
  {C.}~\bibnamefont {Song}},\ }\bibfield  {title} {\enquote {\bibinfo {title}
  {Orthogonal interlayer coupling in an all-antiferromagnetic junction},}\
  }\href {https://doi.org/10.1038/s41467-022-31531-w} {\bibfield  {journal}
  {\bibinfo  {journal} {Nature Communications}\ }\textbf {\bibinfo {volume}
  {13}} (\bibinfo {year} {2022}),\ 10.1038/s41467-022-31531-w}\BibitemShut
  {NoStop}%
\bibitem [{\citenamefont {Prokopenko}\ and\ \citenamefont
  {Prokopenko}(2024)}]{Prokopenko2024}%
  \BibitemOpen
  \bibfield  {author} {\bibinfo {author} {\bibfnamefont {V.}~\bibnamefont
  {Prokopenko}}\ and\ \bibinfo {author} {\bibfnamefont {O.}~\bibnamefont
  {Prokopenko}},\ }\enquote {\bibinfo {title} {Terahertz signal detectors based
  on antiferromagnetic spintronic nanostructures},}\ in\ \href
  {https://doi.org/10.1007/978-981-97-2667-7_4} {\emph {\bibinfo {booktitle}
  {Nanocomposite and Nanocrystalline Materials and Coatings: Microstructure,
  Properties and Applications}}},\ \bibinfo {editor} {edited by\ \bibinfo
  {editor} {\bibfnamefont {A.~D.}\ \bibnamefont {Pogrebnjak}}, \bibinfo
  {editor} {\bibfnamefont {Y.}~\bibnamefont {Bing}},\ and\ \bibinfo {editor}
  {\bibfnamefont {M.}~\bibnamefont {Sahul}}}\ (\bibinfo  {publisher} {Springer
  Nature Singapore},\ \bibinfo {address} {Singapore},\ \bibinfo {year} {2024})\
  pp.\ \bibinfo {pages} {129--147}\BibitemShut {NoStop}%
\bibitem [{\citenamefont {To}\ \emph {et~al.}(2024)\citenamefont {To},
  \citenamefont {Rai}, \citenamefont {Zide}, \citenamefont {Law}, \citenamefont
  {Xiao}, \citenamefont {Jungfleisch},\ and\ \citenamefont {Doty}}]{To2024}%
  \BibitemOpen
  \bibfield  {author} {\bibinfo {author} {\bibfnamefont {D.-Q.}\ \bibnamefont
  {To}}, \bibinfo {author} {\bibfnamefont {A.}~\bibnamefont {Rai}}, \bibinfo
  {author} {\bibfnamefont {J.~M.~O.}\ \bibnamefont {Zide}}, \bibinfo {author}
  {\bibfnamefont {S.}~\bibnamefont {Law}}, \bibinfo {author} {\bibfnamefont
  {J.~Q.}\ \bibnamefont {Xiao}}, \bibinfo {author} {\bibfnamefont {M.~B.}\
  \bibnamefont {Jungfleisch}},\ and\ \bibinfo {author} {\bibfnamefont {M.~F.}\
  \bibnamefont {Doty}},\ }\bibfield  {title} {\enquote {\bibinfo {title}
  {Hybridized magnonic materials for $\mathrm{T}\mathrm{H}\mathrm{z}$ frequency
  applications},}\ }\href {https://doi.org/10.1063/5.0189678} {\bibfield
  {journal} {\bibinfo  {journal} {Applied Physics Letters}\ }\textbf {\bibinfo
  {volume} {124}},\ \bibinfo {pages} {082405} (\bibinfo {year}
  {2024})}\BibitemShut {NoStop}%
\bibitem [{\citenamefont {Kanistras}\ \emph {et~al.}(2024)\citenamefont
  {Kanistras}, \citenamefont {Scheuer}, \citenamefont {Anyfantis},
  \citenamefont {Barnasas}, \citenamefont {Torosyan}, \citenamefont {Beigang},
  \citenamefont {Crisan}, \citenamefont {Poulopoulos},\ and\ \citenamefont
  {Papaioannou}}]{Kanistras2024}%
  \BibitemOpen
  \bibfield  {author} {\bibinfo {author} {\bibfnamefont {N.}~\bibnamefont
  {Kanistras}}, \bibinfo {author} {\bibfnamefont {L.}~\bibnamefont {Scheuer}},
  \bibinfo {author} {\bibfnamefont {D.~I.}\ \bibnamefont {Anyfantis}}, \bibinfo
  {author} {\bibfnamefont {A.}~\bibnamefont {Barnasas}}, \bibinfo {author}
  {\bibfnamefont {G.}~\bibnamefont {Torosyan}}, \bibinfo {author}
  {\bibfnamefont {R.}~\bibnamefont {Beigang}}, \bibinfo {author} {\bibfnamefont
  {O.}~\bibnamefont {Crisan}}, \bibinfo {author} {\bibfnamefont
  {P.}~\bibnamefont {Poulopoulos}},\ and\ \bibinfo {author} {\bibfnamefont
  {E.~T.}\ \bibnamefont {Papaioannou}},\ }\bibfield  {title} {\enquote
  {\bibinfo {title} {Magnetic properties and $\mathrm{T}\mathrm{H}\mathrm{z}$
  emission from
  $\mathrm{C}\mathrm{o}/\mathrm{C}\mathrm{o}\mathrm{O}/\mathrm{P}\mathrm{t}$
  and
  $\mathrm{N}\mathrm{i}/\mathrm{N}\mathrm{i}\mathrm{O}/\mathrm{P}\mathrm{t}$
  trilayers},}\ }\href {https://doi.org/10.3390/nano14020215} {\bibfield
  {journal} {\bibinfo  {journal} {Nanomaterials}\ }\textbf {\bibinfo {volume}
  {14}} (\bibinfo {year} {2024}),\ 10.3390/nano14020215}\BibitemShut {NoStop}%
\bibitem [{\citenamefont {Saito}\ \emph {et~al.}(2023)\citenamefont {Saito},
  \citenamefont {Kholid}, \citenamefont {Karashtin}, \citenamefont
  {Pashenkin},\ and\ \citenamefont {Mikhaylovskiy}}]{Saito2023}%
  \BibitemOpen
  \bibfield  {author} {\bibinfo {author} {\bibfnamefont {Y.}~\bibnamefont
  {Saito}}, \bibinfo {author} {\bibfnamefont {F.~N.}\ \bibnamefont {Kholid}},
  \bibinfo {author} {\bibfnamefont {E.}~\bibnamefont {Karashtin}}, \bibinfo
  {author} {\bibfnamefont {I.}~\bibnamefont {Pashenkin}},\ and\ \bibinfo
  {author} {\bibfnamefont {R.~V.}\ \bibnamefont {Mikhaylovskiy}},\ }\bibfield
  {title} {\enquote {\bibinfo {title} {Terahertz emission spectroscopy of
  exchange-biased spintronic heterostructures: Single- and double-pump
  techniques},}\ }\href {https://doi.org/10.1103/PhysRevApplied.19.064040}
  {\bibfield  {journal} {\bibinfo  {journal} {Phys. Rev. Appl.}\ }\textbf
  {\bibinfo {volume} {19}},\ \bibinfo {pages} {064040} (\bibinfo {year}
  {2023})}\BibitemShut {NoStop}%
\bibitem [{\citenamefont {Gorbatova}\ \emph {et~al.}(2024)\citenamefont
  {Gorbatova}, \citenamefont {Buryakov}, \citenamefont {Avdeev}, \citenamefont
  {Lebedeva}, \citenamefont {Pashen’kin}, \citenamefont {Karashtin},
  \citenamefont {Sapozhnikov},\ and\ \citenamefont {Mishina}}]{Gorbatova2024}%
  \BibitemOpen
  \bibfield  {author} {\bibinfo {author} {\bibfnamefont {A.~V.}\ \bibnamefont
  {Gorbatova}}, \bibinfo {author} {\bibfnamefont {A.~M.}\ \bibnamefont
  {Buryakov}}, \bibinfo {author} {\bibfnamefont {P.~Y.}\ \bibnamefont
  {Avdeev}}, \bibinfo {author} {\bibfnamefont {E.~D.}\ \bibnamefont
  {Lebedeva}}, \bibinfo {author} {\bibfnamefont {I.~Y.}\ \bibnamefont
  {Pashen’kin}}, \bibinfo {author} {\bibfnamefont {E.~A.}\ \bibnamefont
  {Karashtin}}, \bibinfo {author} {\bibfnamefont {M.~V.}\ \bibnamefont
  {Sapozhnikov}},\ and\ \bibinfo {author} {\bibfnamefont {E.~D.}\ \bibnamefont
  {Mishina}},\ }\bibfield  {title} {\enquote {\bibinfo {title} {Effect of
  laser-induced heating of the antiferromagnetic irmn layer on generation of
  terahertz pulses in
  $\mathrm{C}\mathrm{o}/\mathrm{W}\mathrm{S}\mathrm{e}_{2}$-based spintronic
  emitters},}\ }\href {https://doi.org/10.3103/S1541308X24700262} {\bibfield
  {journal} {\bibinfo  {journal} {Physics of Wave Phenomena}\ }\textbf
  {\bibinfo {volume} {32}},\ \bibinfo {pages} {273–279} (\bibinfo {year}
  {2024})}\BibitemShut {NoStop}%
\bibitem [{\citenamefont {Zhuang}\ \emph {et~al.}(2024)\citenamefont {Zhuang},
  \citenamefont {Zhang}, \citenamefont {Zhu}, \citenamefont {Sun},
  \citenamefont {Eom}, \citenamefont {Evans},\ and\ \citenamefont
  {Hu}}]{Zhuang2024}%
  \BibitemOpen
  \bibfield  {author} {\bibinfo {author} {\bibfnamefont {S.}~\bibnamefont
  {Zhuang}}, \bibinfo {author} {\bibfnamefont {X.}~\bibnamefont {Zhang}},
  \bibinfo {author} {\bibfnamefont {Y.}~\bibnamefont {Zhu}}, \bibinfo {author}
  {\bibfnamefont {N.~X.}\ \bibnamefont {Sun}}, \bibinfo {author} {\bibfnamefont
  {C.-B.}\ \bibnamefont {Eom}}, \bibinfo {author} {\bibfnamefont {P.~G.}\
  \bibnamefont {Evans}},\ and\ \bibinfo {author} {\bibfnamefont {J.-M.}\
  \bibnamefont {Hu}},\ }\bibfield  {title} {\enquote {\bibinfo {title} {Hybrid
  magnon-phonon cavity for large-amplitude terahertz spin-wave excitation},}\
  }\href {https://doi.org/10.1103/PhysRevApplied.21.044009} {\bibfield
  {journal} {\bibinfo  {journal} {Phys. Rev. Appl.}\ }\textbf {\bibinfo
  {volume} {21}},\ \bibinfo {pages} {044009} (\bibinfo {year}
  {2024})}\BibitemShut {NoStop}%
\bibitem [{\citenamefont {Gr\"unberg}\ \emph {et~al.}(1986)\citenamefont
  {Gr\"unberg}, \citenamefont {Schreiber}, \citenamefont {Pang}, \citenamefont
  {Brodsky},\ and\ \citenamefont {Sowers}}]{Grunberg1986}%
  \BibitemOpen
  \bibfield  {author} {\bibinfo {author} {\bibfnamefont {P.}~\bibnamefont
  {Gr\"unberg}}, \bibinfo {author} {\bibfnamefont {R.}~\bibnamefont
  {Schreiber}}, \bibinfo {author} {\bibfnamefont {Y.}~\bibnamefont {Pang}},
  \bibinfo {author} {\bibfnamefont {M.~B.}\ \bibnamefont {Brodsky}},\ and\
  \bibinfo {author} {\bibfnamefont {H.}~\bibnamefont {Sowers}},\ }\bibfield
  {title} {\enquote {\bibinfo {title} {Layered magnetic structures: Evidence
  for antiferromagnetic coupling of fe layers across cr interlayers},}\ }\href
  {https://doi.org/10.1103/PhysRevLett.57.2442} {\bibfield  {journal} {\bibinfo
   {journal} {Phys. Rev. Lett.}\ }\textbf {\bibinfo {volume} {57}},\ \bibinfo
  {pages} {2442--2445} (\bibinfo {year} {1986})}\BibitemShut {NoStop}%
\bibitem [{\citenamefont {Cheng}, \citenamefont {Xiao},\ and\ \citenamefont
  {Zhu}(2018{\natexlab{a}})}]{ChengPRL2018}%
  \BibitemOpen
  \bibfield  {author} {\bibinfo {author} {\bibfnamefont {R.}~\bibnamefont
  {Cheng}}, \bibinfo {author} {\bibfnamefont {D.}~\bibnamefont {Xiao}},\ and\
  \bibinfo {author} {\bibfnamefont {J.-G.}\ \bibnamefont {Zhu}},\ }\bibfield
  {title} {\enquote {\bibinfo {title} {Interlayer couplings mediated by
  antiferromagnetic magnons},}\ }\href
  {https://doi.org/10.1103/PhysRevLett.121.207202} {\bibfield  {journal}
  {\bibinfo  {journal} {Phys. Rev. Lett.}\ }\textbf {\bibinfo {volume} {121}},\
  \bibinfo {pages} {207202} (\bibinfo {year} {2018}{\natexlab{a}})}\BibitemShut
  {NoStop}%
\bibitem [{\citenamefont {Cheng}, \citenamefont {Xiao},\ and\ \citenamefont
  {Zhu}(2018{\natexlab{b}})}]{ChengPRB2018}%
  \BibitemOpen
  \bibfield  {author} {\bibinfo {author} {\bibfnamefont {R.}~\bibnamefont
  {Cheng}}, \bibinfo {author} {\bibfnamefont {D.}~\bibnamefont {Xiao}},\ and\
  \bibinfo {author} {\bibfnamefont {J.-G.}\ \bibnamefont {Zhu}},\ }\bibfield
  {title} {\enquote {\bibinfo {title} {Antiferromagnet-based magnonic
  spin-transfer torque},}\ }\href {https://doi.org/10.1103/PhysRevB.98.020408}
  {\bibfield  {journal} {\bibinfo  {journal} {Phys. Rev. B}\ }\textbf {\bibinfo
  {volume} {98}},\ \bibinfo {pages} {020408} (\bibinfo {year}
  {2018}{\natexlab{b}})}\BibitemShut {NoStop}%
\bibitem [{\citenamefont {Hagelschuer}, \citenamefont {Shokr},\ and\
  \citenamefont {Kuch}(2016)}]{Hagelschuer2016}%
  \BibitemOpen
  \bibfield  {author} {\bibinfo {author} {\bibfnamefont {T.}~\bibnamefont
  {Hagelschuer}}, \bibinfo {author} {\bibfnamefont {Y.~A.}\ \bibnamefont
  {Shokr}},\ and\ \bibinfo {author} {\bibfnamefont {W.}~\bibnamefont {Kuch}},\
  }\bibfield  {title} {\enquote {\bibinfo {title} {Spin-state transition in
  antiferromagnetic $\mathrm{N}\mathrm{i}_{0.4}\mathrm{M}\mathrm{n}_{0.6}$
  films in
  $\mathrm{N}\mathrm{i}/\mathrm{N}\mathrm{i}\mathrm{M}\mathrm{n}/\mathrm{N}\mathrm{i}$
  trilayers on $\mathrm{C}\mathrm{u}$(001)},}\ }\href
  {https://doi.org/10.1103/PhysRevB.93.054428} {\bibfield  {journal} {\bibinfo
  {journal} {Phys. Rev. B}\ }\textbf {\bibinfo {volume} {93}},\ \bibinfo
  {pages} {054428} (\bibinfo {year} {2016})}\BibitemShut {NoStop}%
\bibitem [{\citenamefont {Takei}\ \emph {et~al.}(2015)\citenamefont {Takei},
  \citenamefont {Moriyama}, \citenamefont {Ono},\ and\ \citenamefont
  {Tserkovnyak}}]{Takei2015}%
  \BibitemOpen
  \bibfield  {author} {\bibinfo {author} {\bibfnamefont {S.}~\bibnamefont
  {Takei}}, \bibinfo {author} {\bibfnamefont {T.}~\bibnamefont {Moriyama}},
  \bibinfo {author} {\bibfnamefont {T.}~\bibnamefont {Ono}},\ and\ \bibinfo
  {author} {\bibfnamefont {Y.}~\bibnamefont {Tserkovnyak}},\ }\bibfield
  {title} {\enquote {\bibinfo {title} {Antiferromagnet-mediated spin transfer
  between a metal and a ferromagnet},}\ }\href
  {https://doi.org/10.1103/PhysRevB.92.020409} {\bibfield  {journal} {\bibinfo
  {journal} {Phys. Rev. B}\ }\textbf {\bibinfo {volume} {92}},\ \bibinfo
  {pages} {020409} (\bibinfo {year} {2015})}\BibitemShut {NoStop}%
\bibitem [{\citenamefont {Polishchuk}\ \emph {et~al.}(2021)\citenamefont
  {Polishchuk}, \citenamefont {Tykhonenko-Polishchuk}, \citenamefont
  {Lytvynenko}, \citenamefont {Rostas}, \citenamefont {Gomonay},\ and\
  \citenamefont {Korenivski}}]{Polishchuk2021}%
  \BibitemOpen
  \bibfield  {author} {\bibinfo {author} {\bibfnamefont {D.~M.}\ \bibnamefont
  {Polishchuk}}, \bibinfo {author} {\bibfnamefont {Y.~O.}\ \bibnamefont
  {Tykhonenko-Polishchuk}}, \bibinfo {author} {\bibfnamefont {Y.~M.}\
  \bibnamefont {Lytvynenko}}, \bibinfo {author} {\bibfnamefont {A.~M.}\
  \bibnamefont {Rostas}}, \bibinfo {author} {\bibfnamefont {O.~V.}\
  \bibnamefont {Gomonay}},\ and\ \bibinfo {author} {\bibfnamefont
  {V.}~\bibnamefont {Korenivski}},\ }\bibfield  {title} {\enquote {\bibinfo
  {title} {Thermal gating of magnon exchange in magnetic multilayers with
  antiferromagnetic spacers},}\ }\href
  {https://doi.org/10.1103/PhysRevLett.126.227203} {\bibfield  {journal}
  {\bibinfo  {journal} {Phys. Rev. Lett.}\ }\textbf {\bibinfo {volume} {126}},\
  \bibinfo {pages} {227203} (\bibinfo {year} {2021})}\BibitemShut {NoStop}%
\bibitem [{\citenamefont {Polishchuk}\ \emph {et~al.}(2023)\citenamefont
  {Polishchuk}, \citenamefont {Tykhonenko-Polishchuk}, \citenamefont
  {Lytvynenko}, \citenamefont {Rostas}, \citenamefont {Kuncser}, \citenamefont
  {Kravets}, \citenamefont {Tovstolytkin}, \citenamefont {Gomonay},\ and\
  \citenamefont {Korenivski}}]{Polishchuk2023}%
  \BibitemOpen
  \bibfield  {author} {\bibinfo {author} {\bibfnamefont {D.~M.}\ \bibnamefont
  {Polishchuk}}, \bibinfo {author} {\bibfnamefont {Y.~O.}\ \bibnamefont
  {Tykhonenko-Polishchuk}}, \bibinfo {author} {\bibfnamefont {Y.~M.}\
  \bibnamefont {Lytvynenko}}, \bibinfo {author} {\bibfnamefont {A.~M.}\
  \bibnamefont {Rostas}}, \bibinfo {author} {\bibfnamefont {V.}~\bibnamefont
  {Kuncser}}, \bibinfo {author} {\bibfnamefont {A.~F.}\ \bibnamefont
  {Kravets}}, \bibinfo {author} {\bibfnamefont {A.~I.}\ \bibnamefont
  {Tovstolytkin}}, \bibinfo {author} {\bibfnamefont {O.~V.}\ \bibnamefont
  {Gomonay}},\ and\ \bibinfo {author} {\bibfnamefont {V.}~\bibnamefont
  {Korenivski}},\ }\bibfield  {title} {\enquote {\bibinfo {title}
  {Antiferromagnet-mediated interlayer exchange: Hybridization versus proximity
  effect},}\ }\href {https://doi.org/10.1103/PhysRevB.107.224432} {\bibfield
  {journal} {\bibinfo  {journal} {Phys. Rev. B}\ }\textbf {\bibinfo {volume}
  {107}},\ \bibinfo {pages} {224432} (\bibinfo {year} {2023})}\BibitemShut
  {NoStop}%
\bibitem [{\citenamefont {Tobia}\ \emph {et~al.}(2008)\citenamefont {Tobia},
  \citenamefont {Winkler}, \citenamefont {Zysler}, \citenamefont {Granada},\
  and\ \citenamefont {Troiani}}]{Tobia2008}%
  \BibitemOpen
  \bibfield  {author} {\bibinfo {author} {\bibfnamefont {D.}~\bibnamefont
  {Tobia}}, \bibinfo {author} {\bibfnamefont {E.}~\bibnamefont {Winkler}},
  \bibinfo {author} {\bibfnamefont {R.~D.}\ \bibnamefont {Zysler}}, \bibinfo
  {author} {\bibfnamefont {M.}~\bibnamefont {Granada}},\ and\ \bibinfo {author}
  {\bibfnamefont {H.~E.}\ \bibnamefont {Troiani}},\ }\bibfield  {title}
  {\enquote {\bibinfo {title} {Size dependence of the magnetic properties of
  antiferromagnetic $\mathrm{C}\mathrm{r}_{2}\mathrm{O}_{3}$ nanoparticles},}\
  }\href {https://doi.org/10.1103/PhysRevB.78.104412} {\bibfield  {journal}
  {\bibinfo  {journal} {Phys. Rev. B}\ }\textbf {\bibinfo {volume} {78}},\
  \bibinfo {pages} {104412} (\bibinfo {year} {2008})}\BibitemShut {NoStop}%
\bibitem [{\citenamefont {Lenz}, \citenamefont {Zander},\ and\ \citenamefont
  {Kuch}(2007)}]{Lenz2007}%
  \BibitemOpen
  \bibfield  {author} {\bibinfo {author} {\bibfnamefont {K.}~\bibnamefont
  {Lenz}}, \bibinfo {author} {\bibfnamefont {S.}~\bibnamefont {Zander}},\ and\
  \bibinfo {author} {\bibfnamefont {W.}~\bibnamefont {Kuch}},\ }\bibfield
  {title} {\enquote {\bibinfo {title} {Magnetic proximity effects in
  antiferromagnet/ferromagnet bilayers: The impact on the $\mathrm{N}$\'eel
  temperature},}\ }\href {https://doi.org/10.1103/PhysRevLett.98.237201}
  {\bibfield  {journal} {\bibinfo  {journal} {Phys. Rev. Lett.}\ }\textbf
  {\bibinfo {volume} {98}},\ \bibinfo {pages} {237201} (\bibinfo {year}
  {2007})}\BibitemShut {NoStop}%
\bibitem [{\citenamefont {Golosovsky}\ \emph {et~al.}(2009)\citenamefont
  {Golosovsky}, \citenamefont {Salazar-Alvarez}, \citenamefont
  {L\'opez-Ortega}, \citenamefont {Gonz\'alez}, \citenamefont {Sort},
  \citenamefont {Estrader}, \citenamefont {Suri\~nach}, \citenamefont
  {Bar\'o},\ and\ \citenamefont {Nogu\'es}}]{Golosovsky2009}%
  \BibitemOpen
  \bibfield  {author} {\bibinfo {author} {\bibfnamefont {I.~V.}\ \bibnamefont
  {Golosovsky}}, \bibinfo {author} {\bibfnamefont {G.}~\bibnamefont
  {Salazar-Alvarez}}, \bibinfo {author} {\bibfnamefont {A.}~\bibnamefont
  {L\'opez-Ortega}}, \bibinfo {author} {\bibfnamefont {M.~A.}\ \bibnamefont
  {Gonz\'alez}}, \bibinfo {author} {\bibfnamefont {J.}~\bibnamefont {Sort}},
  \bibinfo {author} {\bibfnamefont {M.}~\bibnamefont {Estrader}}, \bibinfo
  {author} {\bibfnamefont {S.}~\bibnamefont {Suri\~nach}}, \bibinfo {author}
  {\bibfnamefont {M.~D.}\ \bibnamefont {Bar\'o}},\ and\ \bibinfo {author}
  {\bibfnamefont {J.}~\bibnamefont {Nogu\'es}},\ }\bibfield  {title} {\enquote
  {\bibinfo {title} {Magnetic proximity effect features in
  antiferromagnetic/ferrimagnetic core-shell nanoparticles},}\ }\href
  {https://doi.org/10.1103/PhysRevLett.102.247201} {\bibfield  {journal}
  {\bibinfo  {journal} {Phys. Rev. Lett.}\ }\textbf {\bibinfo {volume} {102}},\
  \bibinfo {pages} {247201} (\bibinfo {year} {2009})}\BibitemShut {NoStop}%
\bibitem [{\citenamefont {Antel}, \citenamefont {Perjeru},\ and\ \citenamefont
  {Harp}(1999)}]{Antel1999}%
  \BibitemOpen
  \bibfield  {author} {\bibinfo {author} {\bibfnamefont {W.~J.}\ \bibnamefont
  {Antel}}, \bibinfo {author} {\bibfnamefont {F.}~\bibnamefont {Perjeru}},\
  and\ \bibinfo {author} {\bibfnamefont {G.~R.}\ \bibnamefont {Harp}},\
  }\bibfield  {title} {\enquote {\bibinfo {title} {Spin structure at the
  interface of exchange biased
  $\mathrm{F}\mathrm{e}\mathrm{M}\mathrm{n}/\mathrm{C}\mathrm{o}$ bilayers},}\
  }\href {https://doi.org/10.1103/PhysRevLett.83.1439} {\bibfield  {journal}
  {\bibinfo  {journal} {Phys. Rev. Lett.}\ }\textbf {\bibinfo {volume} {83}},\
  \bibinfo {pages} {1439--1442} (\bibinfo {year} {1999})}\BibitemShut {NoStop}%
\bibitem [{\citenamefont {Liu}\ and\ \citenamefont
  {Adenwalla}(2003)}]{Liu2003}%
  \BibitemOpen
  \bibfield  {author} {\bibinfo {author} {\bibfnamefont {Z.~Y.}\ \bibnamefont
  {Liu}}\ and\ \bibinfo {author} {\bibfnamefont {S.}~\bibnamefont
  {Adenwalla}},\ }\bibfield  {title} {\enquote {\bibinfo {title} {Oscillatory
  interlayer exchange coupling and its temperature dependence in
  $[\mathrm{P}\mathrm{t}/\mathrm{C}\mathrm{o}{]}_{3}/\mathrm{N}\mathrm{i}\mathrm{O}/[\mathrm{C}\mathrm{o}/\mathrm{P}\mathrm{t}{]}_{3}$
  multilayers with perpendicular anisotropy},}\ }\href
  {https://doi.org/10.1103/PhysRevLett.91.037207} {\bibfield  {journal}
  {\bibinfo  {journal} {Phys. Rev. Lett.}\ }\textbf {\bibinfo {volume} {91}},\
  \bibinfo {pages} {037207} (\bibinfo {year} {2003})}\BibitemShut {NoStop}%
\bibitem [{\citenamefont {Gomonay}\ \emph {et~al.}(2018)\citenamefont
  {Gomonay}, \citenamefont {Baltz}, \citenamefont {Brataas},\ and\
  \citenamefont {Tserkovnyak}}]{Gomonay2018}%
  \BibitemOpen
  \bibfield  {author} {\bibinfo {author} {\bibfnamefont {O.}~\bibnamefont
  {Gomonay}}, \bibinfo {author} {\bibfnamefont {V.}~\bibnamefont {Baltz}},
  \bibinfo {author} {\bibfnamefont {A.}~\bibnamefont {Brataas}},\ and\ \bibinfo
  {author} {\bibfnamefont {Y.}~\bibnamefont {Tserkovnyak}},\ }\bibfield
  {title} {\enquote {\bibinfo {title} {Antiferromagnetic spin textures and
  dynamics},}\ }\href {https://doi.org/10.1038/s41567-018-0049-4} {\bibfield
  {journal} {\bibinfo  {journal} {Nature Physics}\ }\textbf {\bibinfo {volume}
  {14}} (\bibinfo {year} {2018}),\ 10.1038/s41567-018-0049-4}\BibitemShut
  {NoStop}%
\bibitem [{\citenamefont {Li}\ \emph {et~al.}(2020)\citenamefont {Li},
  \citenamefont {Wilson}, \citenamefont {Cheng}, \citenamefont {Lohmann},
  \citenamefont {Kavand}, \citenamefont {Yuan}, \citenamefont {Aldosary},
  \citenamefont {Agladze}, \citenamefont {Wei}, \citenamefont {Sherwin},\ and\
  \citenamefont {Shi}}]{Li2020}%
  \BibitemOpen
  \bibfield  {author} {\bibinfo {author} {\bibfnamefont {J.}~\bibnamefont
  {Li}}, \bibinfo {author} {\bibfnamefont {C.~B.}\ \bibnamefont {Wilson}},
  \bibinfo {author} {\bibfnamefont {R.}~\bibnamefont {Cheng}}, \bibinfo
  {author} {\bibfnamefont {M.}~\bibnamefont {Lohmann}}, \bibinfo {author}
  {\bibfnamefont {M.}~\bibnamefont {Kavand}}, \bibinfo {author} {\bibfnamefont
  {W.}~\bibnamefont {Yuan}}, \bibinfo {author} {\bibfnamefont {M.}~\bibnamefont
  {Aldosary}}, \bibinfo {author} {\bibfnamefont {N.}~\bibnamefont {Agladze}},
  \bibinfo {author} {\bibfnamefont {P.}~\bibnamefont {Wei}}, \bibinfo {author}
  {\bibfnamefont {M.~S.}\ \bibnamefont {Sherwin}},\ and\ \bibinfo {author}
  {\bibfnamefont {J.}~\bibnamefont {Shi}},\ }\bibfield  {title} {\enquote
  {\bibinfo {title} {Spin current from sub-terahertz-generated
  antiferromagnetic magnons},}\ }\href
  {https://doi.org/10.1038/s41586-020-1950-4} {\bibfield  {journal} {\bibinfo
  {journal} {Nature}\ }\textbf {\bibinfo {volume} {578}} (\bibinfo {year}
  {2020}),\ 10.1038/s41586-020-1950-4}\BibitemShut {NoStop}%
\bibitem [{\citenamefont {Vansteenkiste}\ \emph {et~al.}(2014)\citenamefont
  {Vansteenkiste}, \citenamefont {Leliaert}, \citenamefont {Dvornik},
  \citenamefont {Helsen}, \citenamefont {Garcia-Sanchez},\ and\ \citenamefont
  {Van~Waeyenberge}}]{mumax}%
  \BibitemOpen
  \bibfield  {author} {\bibinfo {author} {\bibfnamefont {A.}~\bibnamefont
  {Vansteenkiste}}, \bibinfo {author} {\bibfnamefont {J.}~\bibnamefont
  {Leliaert}}, \bibinfo {author} {\bibfnamefont {M.}~\bibnamefont {Dvornik}},
  \bibinfo {author} {\bibfnamefont {M.}~\bibnamefont {Helsen}}, \bibinfo
  {author} {\bibfnamefont {F.}~\bibnamefont {Garcia-Sanchez}},\ and\ \bibinfo
  {author} {\bibfnamefont {B.}~\bibnamefont {Van~Waeyenberge}},\ }\bibfield
  {title} {\enquote {\bibinfo {title} {The design and verification of
  mumax3},}\ }\href {https://doi.org/10.1063/1.4899186} {\bibfield  {journal}
  {\bibinfo  {journal} {AIP Advances}\ }\textbf {\bibinfo {volume} {4}},\
  \bibinfo {pages} {107133} (\bibinfo {year} {2014})}\BibitemShut {NoStop}%
\bibitem [{\citenamefont {Demokritov}\ and\ \citenamefont
  {Hillebrands}(2002)}]{Hillebrands2002}%
  \BibitemOpen
  \bibfield  {author} {\bibinfo {author} {\bibfnamefont {S.~O.}\ \bibnamefont
  {Demokritov}}\ and\ \bibinfo {author} {\bibfnamefont {B.}~\bibnamefont
  {Hillebrands}},\ }\enquote {\bibinfo {title} {Spinwaves in laterally confined
  magnetic structures},}\ in\ \href {https://doi.org/10.1007/3-540-40907-6}
  {\emph {\bibinfo {booktitle} {Spin Dynamics in Confined Magnetic Structures
  I}}},\ \bibinfo {editor} {edited by\ \bibinfo {editor} {\bibfnamefont
  {K.~O.}\ \bibnamefont {Burkard~Hillebrands}}}\ (\bibinfo  {publisher}
  {Springer},\ \bibinfo {address} {Berlin},\ \bibinfo {year} {2002})\ pp.\
  \bibinfo {pages} {65--93}\BibitemShut {NoStop}%
\bibitem [{\citenamefont {Vovk}\ \emph {et~al.}(2021)\citenamefont {Vovk},
  \citenamefont {Bunyaev}, \citenamefont {Štrichovanec}, \citenamefont {Vovk},
  \citenamefont {Postolnyi}, \citenamefont {Apolinario}, \citenamefont {Pardo},
  \citenamefont {Algarabel}, \citenamefont {Kakazei},\ and\ \citenamefont
  {Araujo}}]{Vovk2021}%
  \BibitemOpen
  \bibfield  {author} {\bibinfo {author} {\bibfnamefont {A.}~\bibnamefont
  {Vovk}}, \bibinfo {author} {\bibfnamefont {S.~A.}\ \bibnamefont {Bunyaev}},
  \bibinfo {author} {\bibfnamefont {P.}~\bibnamefont {Štrichovanec}}, \bibinfo
  {author} {\bibfnamefont {N.~R.}\ \bibnamefont {Vovk}}, \bibinfo {author}
  {\bibfnamefont {B.}~\bibnamefont {Postolnyi}}, \bibinfo {author}
  {\bibfnamefont {A.}~\bibnamefont {Apolinario}}, \bibinfo {author}
  {\bibfnamefont {J.~n.}\ \bibnamefont {Pardo}}, \bibinfo {author}
  {\bibfnamefont {P.~A.}\ \bibnamefont {Algarabel}}, \bibinfo {author}
  {\bibfnamefont {G.~N.}\ \bibnamefont {Kakazei}},\ and\ \bibinfo {author}
  {\bibfnamefont {J.~P.}\ \bibnamefont {Araujo}},\ }\bibfield  {title}
  {\enquote {\bibinfo {title} {Control of structural and magnetic properties of
  polycrystalline
  $\mathrm{C}\mathrm{o}_{2}\mathrm{F}\mathrm{e}\mathrm{G}\mathrm{e}$ films via
  deposition and annealing temperatures},}\ }\href
  {https://doi.org/10.3390/nano11051229} {\bibfield  {journal} {\bibinfo
  {journal} {Nanomaterials}\ }\textbf {\bibinfo {volume} {11}} (\bibinfo {year}
  {2021}),\ 10.3390/nano11051229}\BibitemShut {NoStop}%
\bibitem [{\citenamefont {Kravets}\ \emph {et~al.}(2012)\citenamefont
  {Kravets}, \citenamefont {Timoshevskii}, \citenamefont {Yanchitsky},
  \citenamefont {Bergmann}, \citenamefont {Buhler}, \citenamefont {Andersson},\
  and\ \citenamefont {Korenivski}}]{kravets2012temperature}%
  \BibitemOpen
  \bibfield  {author} {\bibinfo {author} {\bibfnamefont {A.~F.}\ \bibnamefont
  {Kravets}}, \bibinfo {author} {\bibfnamefont {A.}~\bibnamefont
  {Timoshevskii}}, \bibinfo {author} {\bibfnamefont {B.}~\bibnamefont
  {Yanchitsky}}, \bibinfo {author} {\bibfnamefont {M.~A.}\ \bibnamefont
  {Bergmann}}, \bibinfo {author} {\bibfnamefont {J.}~\bibnamefont {Buhler}},
  \bibinfo {author} {\bibfnamefont {S.}~\bibnamefont {Andersson}},\ and\
  \bibinfo {author} {\bibfnamefont {V.}~\bibnamefont {Korenivski}},\ }\bibfield
   {title} {\enquote {\bibinfo {title} {Temperature-controlled interlayer
  exchange coupling in strong/weak ferromagnetic multilayers: A thermomagnetic
  curie switch},}\ }\href@noop {} {\bibfield  {journal} {\bibinfo  {journal}
  {Physical Review B—Condensed Matter and Materials Physics}\ }\textbf
  {\bibinfo {volume} {86}},\ \bibinfo {pages} {214413} (\bibinfo {year}
  {2012})}\BibitemShut {NoStop}%
\bibitem [{\citenamefont {Kravets}\ \emph {et~al.}(2015)\citenamefont
  {Kravets}, \citenamefont {Tovstolytkin}, \citenamefont {Dzhezherya},
  \citenamefont {Polishchuk}, \citenamefont {Kozak},\ and\ \citenamefont
  {Korenivski}}]{kravets2015spin}%
  \BibitemOpen
  \bibfield  {author} {\bibinfo {author} {\bibfnamefont {A.}~\bibnamefont
  {Kravets}}, \bibinfo {author} {\bibfnamefont {A.}~\bibnamefont
  {Tovstolytkin}}, \bibinfo {author} {\bibfnamefont {Y.~I.}\ \bibnamefont
  {Dzhezherya}}, \bibinfo {author} {\bibfnamefont {D.}~\bibnamefont
  {Polishchuk}}, \bibinfo {author} {\bibfnamefont {I.}~\bibnamefont {Kozak}},\
  and\ \bibinfo {author} {\bibfnamefont {V.}~\bibnamefont {Korenivski}},\
  }\bibfield  {title} {\enquote {\bibinfo {title} {Spin dynamics in a
  curie-switch},}\ }\href@noop {} {\bibfield  {journal} {\bibinfo  {journal}
  {Journal of Physics: Condensed Matter}\ }\textbf {\bibinfo {volume} {27}},\
  \bibinfo {pages} {446003} (\bibinfo {year} {2015})}\BibitemShut {NoStop}%
\bibitem [{\citenamefont {Kravets}\ \emph {et~al.}(2016)\citenamefont
  {Kravets}, \citenamefont {Polishchuk}, \citenamefont {Dzhezherya},
  \citenamefont {Tovstolytkin}, \citenamefont {Golub},\ and\ \citenamefont
  {Korenivski}}]{kravets2016anisotropic}%
  \BibitemOpen
  \bibfield  {author} {\bibinfo {author} {\bibfnamefont {A.}~\bibnamefont
  {Kravets}}, \bibinfo {author} {\bibfnamefont {D.~M.}\ \bibnamefont
  {Polishchuk}}, \bibinfo {author} {\bibfnamefont {Y.~I.}\ \bibnamefont
  {Dzhezherya}}, \bibinfo {author} {\bibfnamefont {A.}~\bibnamefont
  {Tovstolytkin}}, \bibinfo {author} {\bibfnamefont {V.}~\bibnamefont
  {Golub}},\ and\ \bibinfo {author} {\bibfnamefont {V.}~\bibnamefont
  {Korenivski}},\ }\bibfield  {title} {\enquote {\bibinfo {title} {Anisotropic
  magnetization relaxation in ferromagnetic multilayers with variable
  interlayer exchange coupling},}\ }\href@noop {} {\bibfield  {journal}
  {\bibinfo  {journal} {Physical Review B}\ }\textbf {\bibinfo {volume} {94}},\
  \bibinfo {pages} {064429} (\bibinfo {year} {2016})}\BibitemShut {NoStop}%
\bibitem [{\citenamefont {Kittel}(1958)}]{Kittel1958}%
  \BibitemOpen
  \bibfield  {author} {\bibinfo {author} {\bibfnamefont {C.}~\bibnamefont
  {Kittel}},\ }\bibfield  {title} {\enquote {\bibinfo {title} {Excitation of
  spin waves in a ferromagnet by a uniform rf field},}\ }\href@noop {}
  {\bibfield  {journal} {\bibinfo  {journal} {Phys. Rev.}\ }\textbf {\bibinfo
  {volume} {110}},\ \bibinfo {pages} {1295} (\bibinfo {year}
  {1958})}\BibitemShut {NoStop}%
\bibitem [{\citenamefont {Herring}\ and\ \citenamefont
  {Kittel}(1951)}]{Herring1952}%
  \BibitemOpen
  \bibfield  {author} {\bibinfo {author} {\bibfnamefont {C.}~\bibnamefont
  {Herring}}\ and\ \bibinfo {author} {\bibfnamefont {C.}~\bibnamefont
  {Kittel}},\ }\bibfield  {title} {\enquote {\bibinfo {title} {On the theory of
  spin waves in ferromagnetic media},}\ }\href
  {https://doi.org/10.1103/PhysRev.81.869} {\bibfield  {journal} {\bibinfo
  {journal} {Phys. Rev.}\ }\textbf {\bibinfo {volume} {81}},\ \bibinfo {pages}
  {869--880} (\bibinfo {year} {1951})}\BibitemShut {NoStop}%
\bibitem [{\citenamefont {Jorzick}\ \emph {et~al.}(1999)\citenamefont
  {Jorzick}, \citenamefont {Demokritov}, \citenamefont {Mathieu}, \citenamefont
  {Hillebrands}, \citenamefont {Bartenlian}, \citenamefont {Chappert},
  \citenamefont {Rousseaux},\ and\ \citenamefont {Slavin}}]{Jorzick1999}%
  \BibitemOpen
  \bibfield  {author} {\bibinfo {author} {\bibfnamefont {J.}~\bibnamefont
  {Jorzick}}, \bibinfo {author} {\bibfnamefont {S.~O.}\ \bibnamefont
  {Demokritov}}, \bibinfo {author} {\bibfnamefont {C.}~\bibnamefont {Mathieu}},
  \bibinfo {author} {\bibfnamefont {B.}~\bibnamefont {Hillebrands}}, \bibinfo
  {author} {\bibfnamefont {B.}~\bibnamefont {Bartenlian}}, \bibinfo {author}
  {\bibfnamefont {C.}~\bibnamefont {Chappert}}, \bibinfo {author}
  {\bibfnamefont {F.}~\bibnamefont {Rousseaux}},\ and\ \bibinfo {author}
  {\bibfnamefont {A.~N.}\ \bibnamefont {Slavin}},\ }\bibfield  {title}
  {\enquote {\bibinfo {title} {Brillouin light scattering from quantized spin
  waves in micron-size magnetic wires},}\ }\href
  {https://doi.org/10.1103/PhysRevB.60.15194} {\bibfield  {journal} {\bibinfo
  {journal} {Phys. Rev. B}\ }\textbf {\bibinfo {volume} {60}},\ \bibinfo
  {pages} {15194--15200} (\bibinfo {year} {1999})}\BibitemShut {NoStop}%
\bibitem [{\citenamefont {Yuan}\ \emph {et~al.}(2023)\citenamefont {Yuan},
  \citenamefont {Liu}, \citenamefont {Chen}, \citenamefont {Liu}, \citenamefont
  {Lan}, \citenamefont {Yu}, \citenamefont {Wu}, \citenamefont {Yan},
  \citenamefont {Yung}, \citenamefont {Xiao}, \citenamefont {Jiang},\ and\
  \citenamefont {Yu}}]{Yuan2023}%
  \BibitemOpen
  \bibfield  {author} {\bibinfo {author} {\bibfnamefont {S.}~\bibnamefont
  {Yuan}}, \bibinfo {author} {\bibfnamefont {C.}~\bibnamefont {Liu}}, \bibinfo
  {author} {\bibfnamefont {J.}~\bibnamefont {Chen}}, \bibinfo {author}
  {\bibfnamefont {S.}~\bibnamefont {Liu}}, \bibinfo {author} {\bibfnamefont
  {J.}~\bibnamefont {Lan}}, \bibinfo {author} {\bibfnamefont {H.}~\bibnamefont
  {Yu}}, \bibinfo {author} {\bibfnamefont {J.}~\bibnamefont {Wu}}, \bibinfo
  {author} {\bibfnamefont {F.}~\bibnamefont {Yan}}, \bibinfo {author}
  {\bibfnamefont {M.-H.}\ \bibnamefont {Yung}}, \bibinfo {author}
  {\bibfnamefont {J.}~\bibnamefont {Xiao}}, \bibinfo {author} {\bibfnamefont
  {L.}~\bibnamefont {Jiang}},\ and\ \bibinfo {author} {\bibfnamefont
  {D.}~\bibnamefont {Yu}},\ }\bibfield  {title} {\enquote {\bibinfo {title}
  {Spin-wave-based tunable coupler between superconducting flux qubits},}\
  }\href {https://doi.org/10.1103/PhysRevA.107.012434} {\bibfield  {journal}
  {\bibinfo  {journal} {Phys. Rev. A}\ }\textbf {\bibinfo {volume} {107}},\
  \bibinfo {pages} {012434} (\bibinfo {year} {2023})}\BibitemShut {NoStop}%
\end{thebibliography}%

\end{document}